 \documentclass[final,3p,times,twocolumn]{elsarticle}

\usepackage{amssymb, amsmath}
\usepackage{amsthm}

\usepackage{esint}  
\usepackage{braket}  
\usepackage{algorithmic}  
\usepackage{algorithm}  
\usepackage{graphicx}
\usepackage{subfigure}
\usepackage[percent]{overpic}
\usepackage{quantikz}
\usepackage{lineno}

\usepackage[unicode]{hyperref}
\hypersetup{
    colorlinks=true,
    linkcolor=blue,
    urlcolor=blue
}

\journal{Advances in Engineering Software}

\begin{document}

\begin{frontmatter}



\title{A hybrid quantum-classical framework for computational fluid dynamics}

\cortext[cor]{Corresponding author}
 \author[label1]{Chuang-Chao Ye}
 \author[label1]{Ning-Bo An}
 \author[label1]{Teng-Yang Ma}
 \author[label1]{Meng-Han Dou}
 \author[label2]{Wen Bai}
 \author[label3]{Zhao-Yun Chen\corref{cor}}
\ead{chenzhaoyun@iai.ustc.edu.cn}
 \author[label4]{Guo-Ping Guo}
 
 \affiliation[label1]{organization={Origin Quantum Computing Technology (Hefei) Co., Ltd.},
             city={Hefei},
             postcode={230088},
             country={China}}
 \affiliation[label3]{organization={Institute of Artificial Intelligence, Hefei Comprehensive National Science Center},
             city={Hefei},
             postcode={230088},
             country={China}}

 \affiliation[label2]{organization={Chinese Aeronautical Establishment},
             city={Beijing},
             postcode={100012},
             country={China}}    

 \affiliation[label4]{organization={CAS Key Laboratory of Quantum Information, University of Science and Technology of China},
             city={Hefei},
             postcode={230026},
             country={China}}      

\begin{abstract}
Great progress has been made in quantum computing in recent years, providing opportunities to overcome computation resource poverty in many scientific computations like computational fluid dynamics (CFD). In this work, efforts are made to exploit quantum potentialities in CFD, and a hybrid classical and quantum computing CFD framework is proposed to release the power of current quantum computing. In this framework, the traditional CFD solvers are coupled with quantum linear algebra libraries in weak form to achieve collaborative computation between classical and quantum computing. The quantum linear solver provides high-precision solutions and scalable problem sizes for linear systems and is designed to be easily callable for solving linear algebra systems similar to classical linear libraries, thus enabling seamless integration into existing CFD solvers. Some typical cases are performed to validate the feasibility of the proposed framework and the correctness of quantum linear algorithms in CFD. 

\end{abstract}


\begin{keyword}


 Hybrid quantum-classical computing; Computational fluid dynamics; Finite volume method; Quantum linear solver.
\end{keyword}

\end{frontmatter}


\section{Introduction}

Computational fluid dynamics (CFD) is becoming increasingly important due to its outstanding capacity to offer more detailed flow information. However, larger sizes of problems and more complex flows mean more computations~\cite{abe2004surface, avsarkisov2014turbulent, lluesma2018influence, federrath2021sonic}.

Classical computation has brought CFD computations to a golden age with the advent of higher-performance processors, more cores, even GPU heterogeneous computing, and the development of parallel techniques \cite{witherden2014pyfr, economon2016su2}. The performance of classical computing relies on the increment of transistor density of processors. The famous Moore's Law predicts that processor performance doubles approximately every two years. However, classical computing reaches its bottleneck because the smallest unit transistor of a classic processor is now close to the atomic level~\cite{wu2022vertical}, making it increasingly difficult to design more powerful processors. People have to exploit more powerful computational paradigms to satisfy increasing computation needs. Quantum computers, first proposed by Richard Feynman in 1982, perform calculations based on the principle of quantum mechanics. The unique properties of superposition and entanglement of qubits allow quantum computers to perform certain calculations much more efficiently than classical computers. Many algorithms have been designed to solve specific problems efficiently on quantum computers, such as the Shor algorithm~\cite{shor1999polynomial} theoretically achieves an exponential speedup over classical algorithms in breaking RSA encryption.

Linear systems of equations are common in CFD, and solving them is computationally expensive. Recent progress in algorithms for solving linear equations opens the way for CFD simulation. The first quantum linear algorithm is proposed by Harrow, Hassidim, and Lloyd~\cite{harrow2009quantum}, named HHL, which is theoretically proved to achieve exponential acceleration over classical conjugate gradient-type algorithms in solving sparse linear algebra systems. Subsequently, many quantum linear algorithms based on HHL or novel approaches are developed to improve efficiency and precision~\cite{clader2013preconditioned, subacsi2019quantum, gilyen2019quantum, lin2020optimal, costa2022optimal}.
However, when solving larger-scale linear systems, more qubits and deeper quantum circuits are required by the quantum algorithms mentioned above, so these algorithms are not suitable for running large-scale calculation examples on current noisy intermediate-scale quantum (NISQ) equipment.
In order to solve large-scale linear systems on current NISQ devices, variational quantum linear solvers have also been developed~\cite{xu2021variational, patil2022variational, bravo2023variational}, and differential equation solving schemes based on variational quantum methods have been proposed \cite{liu2021variational, li2023variational}.

There have been many attempts to introduce quantum computing into computational fluid dynamics, and the quantum CFD (QCFD) era has begun. Traditional CFD methods have been developed to integrate quantum computing, resulting in the quantum lattice Boltzmann method \cite{todorova2020quantum, sanavio2023quantum, kocherla2024two}, the quantum spectral method~\cite{childs2020quantum}, etc. The novel approach \cite{sedykh2023quantum}  based on machine learning like physics-informed neural networks (PINN) is also involved. Quantum algorithms have been applied in many flow simulations, such as flow over airfoil \cite{chen2022quantum}, Poiseuille and Couette flow~\cite{bharadwaj2023hybrid}, reacting flows~\cite{lu2023quantum}, and the reliability of quantum algorithms in CFD have been proved. However, due to the limitations of classical computing resources and the performance of quantum devices, these methods have not been applied to large-scale engineering simulations. 

Quantum computers and algorithms in the NISQ era can not deal with all the computations in practical applications, at least, not all problems are suitably solved with quantum computers. Hybrid quantum-classical computing is a viable way to utilize quantum computing power in the NISQ era. The variational quantum methods are a type of hybrid quantum-classical approach that has been widely used in solving linear equations and optimization problems. Data conversion efficiency between classical computers and quantum computers, which involves the encoding and decoding mechanisms between binary digits and quantum bits, is important to keep the quantum advantage. Standard quantum state tomography is resource-expensive for data readout. Chen et.al~\cite{chen2022quantum} proposed a quantum method for accelerating finite volume methods using classical input and output, and sublinear acceleration is obtained to keep the quantum advantage. However, this method relies highly on quantum random accessing memory (QRAM), which will not be practical in the near future. Subsequently, they proposed a sparse quantum state tomography method to read out quantum information efficiently on near-term quantum devices~\cite{chen2024enabling}. 


This article presents a CFD framework that enables large-scale engineering fluid simulation with hybrid quantum-classical computing. To apply quantum algorithms, firstly, the governing equations of fluid dynamics are recast as a linear system of equations. Then a quantum linear solver proposed in our previous work is used to solve the large linear equations. The solver can match the sizes of linear systems to various quantum computing resources and provide high-precision solutions. This framework can be easily integrated into many current CFD solvers and is compatible with future quantum computers. It can regarded as a reference for designing hybrid quantum-classical computing applications.

This paper is organized as follows. Firstly, the fluid governing equations and the numerical methods to solve the equations are shown in section \ref{Governing-equations} and \ref{Classical-numerical-algorithms}. Section \ref{Quantum-algorithm} gives details of the quantum algorithms involved in this work. Then in section \ref{pattern}, hybrid quantum-classical architecture and algorithm are introduced, and the feasibility is verified in section \ref{NumExp} through numerical experiments. Finally, in section \ref{ConclusionandOutlook}, the conclusion and outlook are given.

\section{Governing equations} \label{Governing-equations}

The compressible flow is governed by conversation laws as follows
\begin{equation}
\begin{aligned}
  &\dfrac{\partial \rho}{\partial t}+\nabla\cdot (\rho u) = 0 \\
  &\dfrac{\partial(\rho\mathbf{u})}{\partial t}+\nabla\cdot(\rho\mathbf{u}\otimes\mathbf{u}) = -\nabla p+\nabla\cdot\tau \\
  &\frac{\partial(\rho E)}{\partial t}+\nabla\cdot(\rho\mathbf{u}E) =- \nabla\cdot(\mathbf{q}+\tau\cdot\mathbf{u}),
\end{aligned}
\end{equation}
where $\rho$ is the density, $\mathbf{u}$ is the velocity vector, $E$ is the energy, $\tau$ is the viscous stress tensor and $\mathbf{q}$ is the heat flux. The equations above can be transformed into the integral form
\begin{equation} \label{eq:CE-1}
\frac{\partial}{\partial t } \iiint_V \vec{U} dV + \iiint_{V} (\vec{F}_{inv} - \vec{F}_{vis}) dV = 0 ,
\end{equation}
where $\vec{U}$ is the vector of conserved variables in the finite control volume $V$ 
\begin{equation} 
\vec{U} =  \begin{bmatrix}
\rho, & \rho u, & \rho v, & \rho w, & \rho E
\end{bmatrix}^T.
\label{eq:CE-2}
\end{equation}
The inviscid flux vector $\vec{F}_{inv}$ is
\begin{equation} 
\vec{F}_{inv} =  \begin{bmatrix}
\rho u \\ \rho u^2 + p \\ \rho vu \\ \rho wu \\ \rho Eu + pu
\end{bmatrix} \vec{i} +
 \begin{bmatrix}
\rho v \\ \rho uv \\ \rho v^2 + p \\ \rho wv \\ \rho Ev + pv
\end{bmatrix} \vec{j} +
\begin{bmatrix}
\rho w \\ \rho uw \\ \rho vw \\ \rho w^2 + p \\ \rho Ew + pw
\end{bmatrix} \vec{k}.
\label{eq:CE-3}
\end{equation}
The viscous flux vector $\vec{F}_{vis}$ is
\begin{equation} 
\begin{aligned}
\vec{F}_{vis} = 
&\begin{bmatrix} 0 \\ \tau_{xx} \\ \tau_{yx} \\ \tau_{zx} \\ \tau_{xx} u + \tau_{xy} v + \tau_{xz} w -q_{x} \end{bmatrix} \vec{i} + \\
&\begin{bmatrix} 0 \\ \tau_{xy} \\ \tau_{yy} \\ \tau_{zy} \\ \tau_{yx} u + \tau_{yy} v + \tau_{yz} w -q_{y} \end{bmatrix} \vec{j} + \\
&\begin{bmatrix} 0 \\ \tau_{xz} \\ \tau_{yz} \\ \tau_{zz} \\ \tau_{zx} u + \tau_{zy} v + \tau_{zz} w -q_{z} \end{bmatrix} \vec{k}.
\label{eq:CE-3}
\end{aligned}
\end{equation}

\section{Numerical methods} \label{Classical-numerical-algorithms}
\subsection{Discrete method}
The finite volume method(FVM) \cite{economon2016su2} has inherent conservation properties and has good robustness and flexibility in dealing with complex shapes and complex flow problems. It is a reliable method widely used in engineering CFD applications.

Assume that the computational domain $\Omega$ is divided into $N$ non-overlapping elements $V_j, j \in [1, M]$. By applying the divergence theorem to the second term on the left-hand side of the equation (\ref{eq:CE-1}), the volume integral can be transformed to the surface integral, and a semi-discretized equation is obtained
\begin{equation} \label{eq:CNA-1} 
\frac{\partial}{\partial t } \iiint_{V_j} \vec{U} dV + R_j(\vec{U}) = 0  ,
\end{equation}
where 
\begin{equation} 
\begin{aligned}
R_j(\vec{U}) & = \oiint_{S} (\vec{F}_{inv} - \vec{F}_{vis}) \cdot \vec{n} dS \\
& = \sum_{k \in N(j)}(\vec{F}^{inv}_{jk} - \vec{F}^{vis}_{jk}) \Delta S_{jk}.
\end{aligned}
\end{equation}

The temporal term in equation (\ref{eq:CNA-1}) can be discretized with implicit and explicit methods. For explicit methods, the solution of the next time step is computed with the previous information.  To keep numerical stability, the step size of explicit methods is limited by the Courant-Friedrichs-Lewy (CFL) number, making it inefficient for long-time simulations. In contrast, implicit methods are not strictly constrained by the CFL number and a larger step size can be used, which makes it popular in industry applications. The temporal term in equation (\ref{eq:CNA-1}) can be implicitly discretized as follows
\begin{equation} \label{eq:CNA-2}
V_j \frac{\Delta \vec{U}^n}{\Delta t } + \vec{R}_j(\vec{U}^{n+1}) = 0  ,
\end{equation}
where $\Delta \vec{U}^n = \vec{U}^{n+1} - \vec{U}^{n}$. And $\vec{R}(\vec{U}^{n+1})$ can be approximated with the Taylor series around $\vec{R}(\vec{U}^{n})$ 
\begin{equation} \label{eq:CNA-3}
 \vec{R}(\vec{U}^{n+1}) =  \vec{R}(\vec{U}^{n}) + \left(\frac{\partial \vec{R}}{\partial \vec{U} }\right)^n \left(\frac{\partial \vec{U}}{\partial t }\right)^n \Delta t + O(\Delta t^2),
\end{equation}
where $J(\vec{U}):= \frac{\partial \vec{R}}{\partial \vec{U} }$ is the flux Jacobian matrix of the residual vector $\vec{R}$. The temporal term $\frac{\partial \vec{U}}{\partial t }$ is approximated as
\begin{equation} \label{eq:CNA-4}
 \frac{\partial \vec{U}}{\partial t } \approx  \frac{\Delta \vec{U}}{\Delta t }.
\end{equation}
Substitute equation (\ref{eq:CNA-4}) into equation (\ref{eq:CNA-3}) and remove higher-order terms, and then equation (\ref{eq:CNA-3}) can be approximated as
\begin{equation} \label{eq:CNA-5}
 \vec{R}(\vec{U}^{n+1}) \approx  \vec{R}(\vec{U}^n)  + \left(\frac{\partial \vec{R}}{\partial \vec{U} }\right)^n \Delta \vec{U}^n.
\end{equation}
Substitute the resulting linearization back to equation (\ref{eq:CNA-2}), we get the following implicit scheme
\begin{equation} 
\left[\frac{V_j}{\Delta t} I + J(\vec{U}^n) \right]\Delta \vec{U}^n = - \vec{R}(\vec{U}^n) .
\label{eq:CNA-6}
\end{equation}
Now, the partial differential governing equations are recast as a large sparse linear system of equations. This linear system can be solved with any linear algorithm.

\subsection{Methods for solving linear systems of equations}

Linear systems of equations can be solved with direct methods and iterative methods. Direct methods typically involve algebraic manipulations and transformations of the original system of equations to simplify and solve for unknown variables. Direct methods have high accuracy and good stability, but the time complexity and memory consumed increase rapidly as the scale of linear systems. So they are preferred when the system size is relatively small or when a high level of accuracy is required. Iterative methods approximate the solution to a system of linear equations by iteratively improving an initial guess until converges to the true solution at a certain precision. Iterative methods, like CG and GMRES method~\cite{saad2003iterative}, can solve linear systems of equations efficiently with fewer computer resources, especially for large-scale asymmetric sparse linear equations. GMRES employs the Arnoldi process to build an orthonormal basis for the Krylov subspace. This process is a key component in transforming the problem into a smaller Hessenberg matrix problem, which can be solved easily and accurately. The convergence performance of iterative methods is affected by the stiffness of the linear system, which is measured by the condition number of the coefficient matrix. Preconditioning techniques are needed to improve the convergence of solving ill-conditioned linear problems.

Similar to classical linear algorithms, there are two main branches of quantum linear algorithms: direct methods and iterative methods. The HHL algorithm~\cite{harrow2009quantum}, the CKS algorithm~\cite{childs2017quantum}, the QSVT~\cite{gilyen2019quantum}, and the quantum discrete adiabatic linear solution algorithm~\cite{costa2022optimal} are famous direct methods. When the matrix of the linear system is sparse, the quantum direct methods have an exponential acceleration over their classical counterparts in the matrix dimension. Direct quantum methods are one-shot methods, that is, when the circuit completes, the solution is solved. Such circuits are typically broad and deep and need a large number of high-fidelity qubits, especially for large-scale problems. Iterative quantum linear algorithms find solutions like their classical counterparts. A typical class of iterative quantum linear algorithms is the variational quantum linear algorithm (VQLA)~\cite{xu2021variational, bravo2023variational}. It is a hybrid algorithm that relies on optimization algorithms performed on classical computers to optimize the parameters of quantum circuits. This class of algorithms shows good properties of noise resilience, making it particularly promising in the context of NISQ devices. An iterative quantum linear solver was developed in Ref.~\cite{chen2024enabling} and successfully achieved high-precision solutions for linear systems in fluid dynamics on a noisy superconducting quantum computer.

\section{Quantum algorithms} \label{Quantum-algorithm}

\subsection{Quantum state preparation}
Quantum state preparation (QSP) plays the role of preparing data for quantum algorithms, and it is sometimes called initial state preparation. The efficiency of QSP is crucial for quantum computing to keep its advantage over classical computing. There are many ways for initial state preparation~\cite{jian2021quantum}. Amplitude encoding is widely used in practice.

For a set of real numbers $\vec{d} = \{d_j| j \in [N] \},[N]=\{0,1,...,N-1\}$, its amplitude encoding state is
\begin{equation}
\ket{\phi }:=  \frac{1}{D}  \sum_{j=0}^{N-1} d_{j} \ket{j} ,
\label{eq:AC-1}
\end{equation}
where $D$ is the 2-norm of $\vec{d}$. When $\vec{d}$ has no prior information, take $n = \lceil \log_2 N \rceil $ qubits. If $\log_2 N$ is not an integer, round it up and fill the excess with 0.
Algorithm \ref{alg:AC1} shows an amplitude encoding algorithm with complexity $O(N)$. Data sets with specific distributions can be encoded more effectively~\cite{grover2002creating}. Data unsatisfied with this distribution can be converted with the quantum random access memory (QRAM)~\cite{xue2021quantum, chen2022quantum}, and the implementation can be found in Ref.~\cite{grover2002creating}. Notice that QRAM is impractical in near term of quantum computing.

\begin{algorithm}[!h]
	\caption{Amplitude encoding algorithm}
    \label{alg:AC1}
    \renewcommand{\algorithmicrequire}{\textbf{Input:}}
    \renewcommand{\algorithmicensure}{\textbf{Output:}}
    \begin{algorithmic}[1]
        \REQUIRE $\vec{d}$  
 	\ENSURE Amplitude encoding state $\ket{\phi} of \vec{d}$    
        
        \FOR{each $i \in [0,N-1]$}
		  \STATE Initialize $\lceil \log_2 N \rceil $ qubits;
		  \STATE Calculate rotation angle   $$\theta_{i,j} = 2\arccos(r_{i,j}/r_{i+1,2j}), j = 0,1,\ldots, 2^i -1 ,$$
        where $$r_{i,j} = \sqrt{\sum_{k = j 2^{n-i}}^{(j+1) 2^{n-i} - 1} d_k^2};$$
   	\STATE  All real and virtual control combinations of Qubits ($0$ to $i-1$) are traversed, while  $2^i$ controlled $RY(\theta)$ are performed on qubit $i$ in sequence according to rotation angles $\theta_{i,0},\theta_{i,1}, \ldots, \theta_{i,2^i -1}$;
        \ENDFOR
        
	\RETURN  The quantum circuit of amplitude encoding and $\ket{\phi}$.
    \end{algorithmic}
\end{algorithm}

\subsection{Quantum state readout}
In quantum state preparation, classical information is encoded in the states of qubits, and then this quantum system evolves after a series of quantum operations are performed on qubits to undergo quantum algorithms. The results of quantum computations are encoded in quantum states, so they must be extracted back to  classical information that can interpreted by classical systems or users, and this process is called quantum state readout.

Quantum state readout needs quantum measurements. There are several methods and techniques to perform this task, such as projective measurement~\cite{johansen2007projective} and quantum tomography~\cite{kerenidis2019quantum}, and each with its advantages and challenges. The choice of readout method depends on the specific quantum system, the desired accuracy, and the available technology. Quantum state tomography is one of the popular quantum readout methods. 

It reconstructs the full quantum state by performing a series of measurements on different bases. The measurement probabilities are used to reconstruct the density matrix expressed as follows
\begin{equation}
    \rho=\frac12\left(I+\vec{r}\cdot\vec{\sigma}\right),
\end{equation}
where $\vec{r}=(r_x,r_y,r_z)$ is the Bloch vector and $\vec{\sigma}=(\sigma_x,\sigma_y,\sigma_z)$ are the Pauli matrices. For an $n$-qubit system, the density matrix has $2^n \times 2^n$ elements. 

Extracting all information from a quantum system with a large number of qubits is computationally expensive, because the quantum state collapses to one of the basis states when it is measured, and the outcome is a classical bit (either 0 or 1). To estimate the probabilities for basis, it is necessary to measure the qubit repeatedly. For example, measure a qubit 10000 times and get 7000 outcomes of ``0'' and 3000 outcomes of ``1'', then the estimated probabilities for $\ket{0}$ and $\ket{1}$ are 0.7 and 0.3 respectively. 

$L_2$-norm and $L_{\infty}$-norm quantum tomography are two common tomography methods. They differ in their approach to minimizing the error between measured and theoretical probabilities. $L_2$-norm tomography minimizes the overall (average) error, making it suitable for noisy data, while $L_{\infty}$-norm tomography minimizes the maximum error, ensuring the worst-case scenario is tightly controlled. $L_{\infty}$-norm tomography is more computationally expensive than $L_2$-norm tomography. The choice between these methods depends on the requirements and constraints of the problems. A sparse tomography method proposed in Ref.~\cite{chen2024enabling}, an improved $L_{\infty}$-norm quantum tomography, is more efficient for specific data.

\subsection{HHL algorithm}
For a linear system of equations $A \vec{x}=\vec{b}$, where $A$ is an $N \times N$ Hermitian matrix, and  $\vec{b} = \sum_{j=0}^{N-1}\beta_{j} \vec{u}_{j}$ is a normalized vector, the solution to this problem can be expressed with its eigen information as follows
\begin{equation} \label{eqn:hhl0}
    \vec{x} = \sum_{j=0}^{N-1}\beta_{j}\lambda_{j}^{-1} \vec{u}_{j},
\end{equation}
where $\vec{u_{j}}$ and $\lambda_{j}$ are the eigenvector and eigenvalue of matrix $A$ respectively. 

	\begin{figure*}[htbp]
		\centering								
		 \resizebox{0.8\textwidth}{!}{\includegraphics{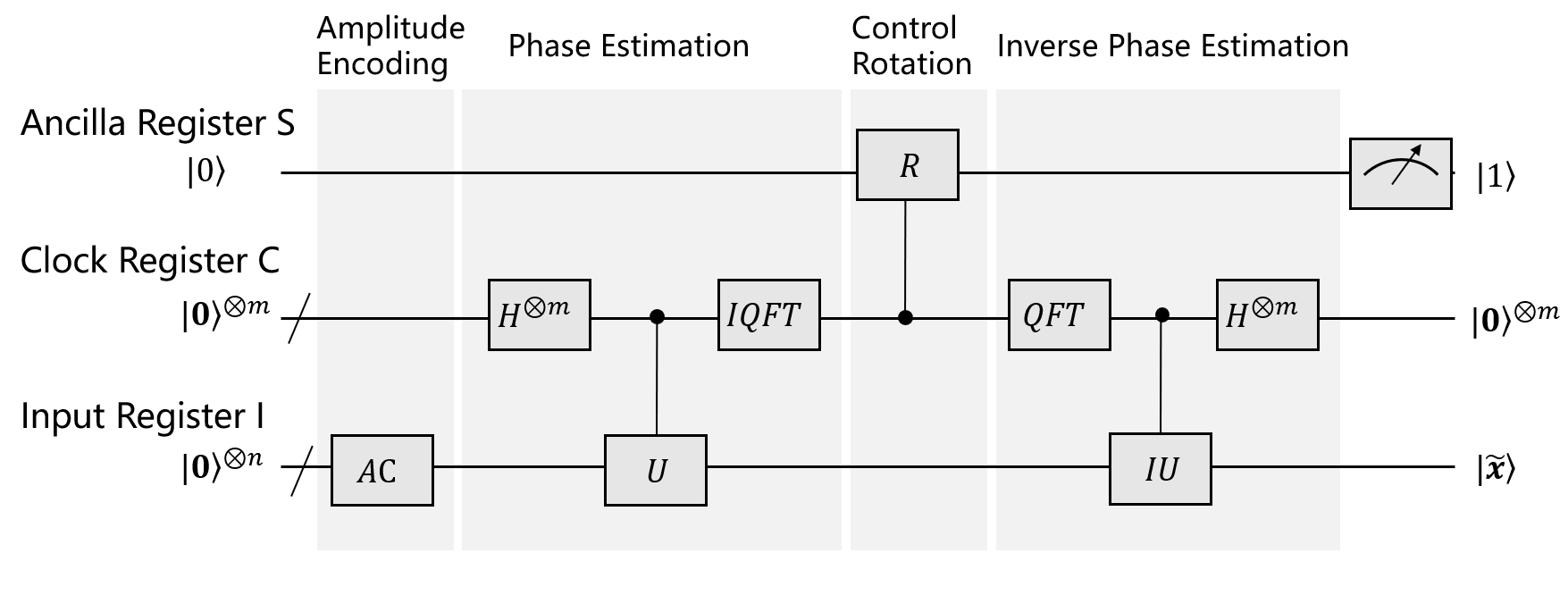}}		
		\caption{HHL algorithm quantum circuit diagram. The HHL algorithm requires three registers. From top to bottom: Ancilla register S, Clock register C, and Input register I.}						
	\label{HHL}
	\end{figure*}
 
In quantum computing, the HHL algorithm~\cite{harrow2009quantum} constructs the solution form above by extracting the eigen information of matrix $A$ with the quantum phase estimation (QPE). Figure \ref{HHL} shows the quantum circuit diagram of the HHL algorithm. To solve the linear problem above, the information of $A$ and $b$ must be encoded into a quantum state. Firstly, $\vec{b}$ is encoded into Input Register I with amplitude encoding as follows
\begin{equation}
\begin{split}
(I\otimes I^{\otimes m} \otimes AC )\ket{0}  \ket{0}^{\otimes m} \ket{0}^{\otimes n} 
&=  \ket{0} \ket{0}^{\otimes m} AC \ket{0}^{\otimes n}\\ 
&= \ket{0} \ket{0}^{\otimes m}  \ket{b},
\end{split}
\label{eq:A-1}
\end{equation}
where $\ket{b} = \sum_{j=0}^{N-1} b_j \ket{j} = \sum_{j=0}^{N-1}\beta_{j} \ket{u_j}$. Matrix $A$ is encoded through Hamiltonian simulation~\cite{berry2007efficient, childs2010relationship} in the following quantum phase estimation (QPE) procedure
\begin{equation}
(I\otimes QPE )\ket{0}  \ket{0}^{\otimes m} \ket{b} = \ket{0} \sum_{j=0}^{N-1} \sum_{k=0}^{T-1} \alpha_{k|j} \beta_{j} \ket{\tilde{\lambda}_k} \ket{u_j},
\label{eq:A-2}
\end{equation}
where $\tilde{\lambda}_k:= 2 \pi k / t_0$. The unitary matrix $U$ in QPE equals  $e^{iAt}$ (here $i^2 =-1$), of which the eigenvalue of $U$ under the eigenvector $\ket{u_j}$ is $e^{i \lambda_jt}$. The eigenvalue can be extracted into the digital base in the form of a superposition state through the inverse quantum Fourier transform~\cite{nielsen2010quantum}.

The eigenvalues have to be extracted onto the amplitude to construct the solution form in equation~(\ref{eqn:hhl0}). To achieve this, an extra ancillary qubit is needed, and rotation operations are performed on it on the condition of qubits carrying 
 information of $\ket{\tilde{\lambda}_k}$ as follows
\begin{equation}
\sum_{j=0}^{N-1} \sum_{k=0}^{T-1}  \left(\sqrt{1-\frac{C^2}{ \tilde{\lambda}^2_k} } \ket{0} + \frac{ C }{ \tilde{\lambda}_k} \ket{1} \right)  \alpha_{k|j} \beta_{j}  \ket{\tilde{\lambda}_k} \ket{u_j},
\label{eq:A-3}
\end{equation}
where $C$ is a value of magnitude $O(1/ \kappa)$. The conditional control rotating gates convert the eigenvalue on a digital basis into the analog state.

To cancel out the information of the eigenvalue on the computational basis,  $\ket{\tilde{\lambda}_k}$ is then uncomputed, thus the quantum state is evolved into 
\begin{equation}
\sum_{j=0}^{N-1} \sum_{k=0}^{T-1}  \left(\sqrt{1-\frac{C^2}{ \tilde{\lambda}^2_k} } \ket{0} + \frac{ C }{ \tilde{\lambda}_k} \ket{1}\right)  \alpha_{k|j} \beta_{j}  \ket{0}^{\otimes m} \ket{u_j}.
\end{equation}

Assuming that the phase estimate is accurate, if $\tilde{\lambda}_k = \lambda_j$, there was $\alpha_{k|j} = 1$, otherwise $\alpha_{k|j} = 0$. So by inverse QPE (IQPE) process, the quantum state of this computing system evolves into
\begin{equation}
\sum_{j=0}^{N-1}  \beta_{j}  \left(\sqrt{1-\frac{C^2}{\lambda^2_j} } \ket{0} + \frac{ C }{ \lambda_j} \ket{1}\right)  \ket{0}^{\otimes m} \ket{u_j}.
\label{eq:A-4}
\end{equation}
When ``1'' is measured at the ancillary qubit, the quantum state of the linear system solution on the output register
\begin{equation}
\ket{\tilde{x}}:= \sqrt{ \frac{1}{\sum_{j=0}^{N-1}C^2 |\beta_j|^2 / |\lambda_j|^2}}\sum_{j=0}^{N-1}  \frac{ C }{ \lambda_j}   \beta_{j} \ket{u_j}
\label{eq:A-5}
\end{equation}
is exactly the normalized form of the solution $\ket{x} = \sum_{j=0}^{N-1}  \frac{\beta_{j} }{ \lambda_j} \ket{u_j}$, which can be extracted using efficient quantum state tomography methods. The final solution $\vec{x}$ of original equations is related with the extracted normalized solution $\vec{\tilde{x}}$ as $\vec{x} = \eta \vec{\tilde{x}}$. The normalization factor $\eta = \sqrt{P_1} \Vert \vec{b} \Vert_2 / C$, where $P_1$ is the probability of measuring ``1'' in the ancillary qubit. For more details on the derivation of the HHL algorithm, readers can refer to Ref.~\cite{zaman2023step,harrow2009quantum}. 

The time complexity of the HHL algorithm is $O(\log(N)s^2 \kappa^2 / \epsilon)$, and it is exponentially faster than the best classical linear algorithm. However, it needs at least $\lceil \log_2 N \rceil + \lceil \log_2 \kappa \rceil + 1$ qubits and the circuit is so deep that it needs much longer coherence time. 


\subsection{Variational quantum linear algorithm}
The variational quantum algorithm~\cite{cerezo2021variational} is a hybrid quantum-classical iterative method with variable-depth quantum circuits. It is regarded as the most promising linear algorithm in the context of current NISQ devices. 

The core idea of the variational quantum linear solver (VQLS) is to construct a Hamiltonian $H = A^{\dagger}(I - \ket{b} \bra{b})A$ that carries the information of the linear system $A \vec{x} = \vec{b}$, and then the ground state of this Hamiltonian quantum is exactly the normalized solution of the linear system of equations. 

\begin{figure*}[htbp]
		\centering								
		\includegraphics[scale=0.9]{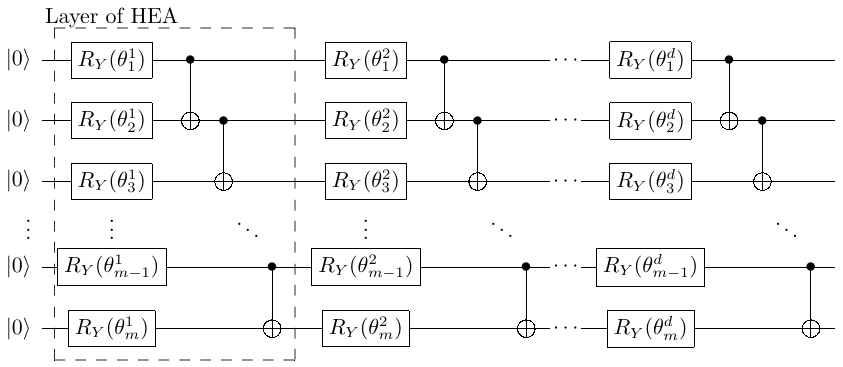}		
		\caption{The circuit of the hardware efficient ansatz.}						
	\label{HEAnsatz}
	\end{figure*}

To start the solver, the vector $\vec{b}$ is directly encoded into quantum circuits, whereas the matrix $A$ is decomposed into a linear weighted summation of a series of unitary matrices, $A = \sum_s^S l_s \sigma_s$, where $l_s$ is a complex scalar coefficient and $\sigma_s$ represents unitary matrix. This decomposition leverages the fundamental properties of quantum mechanics to efficiently manipulate and process the matrix $A$ within the quantum computational framework. To find the ground state of the Hamiltonian, a parameterized quantum state called ansatz, which takes the form  $U_d(\theta_d) \cdots U_1(\theta_1)$, is designed to approximate the solution. The quality of the ansatz is crucial for the convergence and accuracy of the computation. Figure~\ref{HEAnsatz} shows the circuit of the Hardware-efficient ansatz. It is widely used in VQLS due to its flexible layer adjustment.

To quantify the difference between the current quantum state and the desired solution state, a loss function (also called cost function) based on the ansatz and Hamiltonian is constructed as follows
\begin{equation} \label{eq:costfun}
\begin{aligned}
  C(\vec{\theta}) & = \bra{\psi(\vec{\theta})} H \ket{\psi(\vec{\theta})} \\
                  & = \bra{\psi(\vec{\theta})}  A^{\dagger} A \ket{\psi(\vec{\theta})} - \bra{\psi(\vec{\theta})}A^{\dagger}\ket{b}\bra{b}A \ket{\psi(\vec{\theta})}  ,
\end{aligned} 
\end{equation}
where $\ket{\psi(\vec{\theta})}  = U_d(\theta_d) \cdots U_1(\theta_1) \ket{0}$). The gradient descent method is used to update parameters
\begin{equation}
\vec{\theta}_{t+1} = \vec{\theta}_{t} - \beta \nabla C(\vec{\theta}_t) ,
\label{eq:VQLS-3}
\end{equation}
where $\beta$ is the learning rate, and $\nabla C(\vec{\theta}) = \left[\frac{\partial C(\vec{\theta})}{\partial \theta_{1}}, \cdots ,\frac{\partial C(\vec{\theta})}{\partial \theta_{m}}\right]^T$ is the gradient vector at time $t$, where
\begin{equation}
\frac{\partial C(\vec{\theta})}{\partial \theta_{i}} = \frac{\partial (\bra{\psi(\vec{\theta})}  A^{\dagger} A \ket{\psi(\vec{\theta})} - \bra{\psi(\vec{\theta})}A^{\dagger}\ket{b}\bra{b}A \ket{\psi(\vec{\theta})})}{\partial \theta_i }.
\label{eq:VQLS-4}
\end{equation}
 The gradient of the loss function $\nabla C(\vec{\theta})$ can be analytically expressed as
\begin{equation}
\frac{\partial C(\vec{\theta})}{\partial \theta_{i}} = 2\Re \left(\frac{\partial \bra{\psi}}{\partial \theta_i} A^{\dagger}A \ket{\psi} \right)-2\Re \left(\frac{\partial \bra{\psi}}{\partial \theta_i}A^{\dagger}\ket{b}\bra{b}A  \ket{\psi}\right),
\label{eq:VQLS-5}
\end{equation}
where $\Re(\cdot)$ represents the real part, $\frac{\partial \bra{\psi}}{\partial \theta_i} A^{\dagger}A \ket{\psi}$, $\frac{\partial \bra{\psi}}{\partial \theta_i}A^{\dagger} \ket{b}$, and $\bra{b}A \ket{\psi}$ can be obtained by measuring the corresponding quantum circuits.The derivatives of the ansatz can be expressed as
\begin{equation}
\frac{\partial \ket{\psi}}{\partial \theta_i} = \sum_n w_i^{n}\ket{\psi_i^n},
\label{eq:VQLS-6}
\end{equation}
where $w_i^{n}$ is complex coefficient, and
\begin{equation}
\ket{\psi_i^n} = U_d(\theta_d) \cdots \sigma_i^n U_i(\theta_i) \cdots U_1(\theta_1) \ket{0}.
\label{eq:VQLS-7}
\end{equation}
Here $\sigma_i^n$ is obtained through the differential of $U_i(\theta_i)$. Therefore, terms in equation~(\ref{eq:VQLS-5}) can be computed by
\begin{equation}
\frac{\partial \bra{\psi}}{\partial \theta_i} A^{\dagger}A \ket{\psi} = \sum_s \sum_{s'} l_s l_s' \frac{\partial \bra{\psi}}{\partial \theta_i} \sigma_s^{\dagger} \sigma_{s'} \ket{\psi},
\label{eq:VQLS-8}
\end{equation}

\begin{equation}
\frac{\partial \bra{\psi}}{\partial \theta_i}A^{\dagger} \ket{b} = \sum_s  l_s \frac{\partial \bra{\psi}}{\partial \theta_i} \sigma_s^{\dagger}\ket{b} ,
\label{eq:VQLS-9}
\end{equation}
and
\begin{equation}
\bra{b}A \ket{\psi} = \sum_s  l_s \bra{b} \sigma_s^{\dagger}\ket{\psi} .
\label{eq:VQLS-10}
\end{equation}
The corresponding quantum circuits can be found in Ref.~\cite{xu2021variational}. The gradient of the cost function can also be computed with the finite difference method as follows
\begin{equation}
\frac{\partial C(\vec{\theta})}{\partial \theta_{i}} \approx \frac{ C(\vec{\theta} + \Delta \theta_i) - C(\vec{\theta} - \Delta \theta_i) }{2\Delta \theta_{i}}.
\label{eq:VQLS-12}
\end{equation}
The cost value can be computed easily according to equation~(\ref{eq:costfun}), and the quantum states can be extracted efficiently with the sparse tomography method~\cite{chen2024enabling}. Algorithm~\ref{alg:AC2} shows the overall procedure of VQLS.

The algorithm above may suffer convergence difficulty due to the stiffness of the equation and poorly designed anzats. The original problem can be transformed into a new one with the Hamiltonian Morphing method~\cite{subacsi2019quantum} so that it is easier to solve. The idea of Hamiltonian Morphing is to construct a matrix that evolves in time
\begin{equation}
A(\tau) = (1-\tau)I + \tau A ,
\label{eq:VQLS-11}
\end{equation}
where the virtual time $\tau \in [0,1]$. Starting from an identity matrix, $A(\tau)$ approaches the original $A$ as $\tau$ approaches 1 with a small step size. At each time step, the problem $H(\tau) = A^{\dagger}(I - \ket{b}\bra{b})A(\tau)$ is solved, and the parameters $\vec{\theta}(\tau)$ for this resulting solution are taken as an initial value for the problem in the next virtual time step. 
To ensure that $A(\tau)$ is always invertible during the evolution, the problem should be further modified as
\begin{equation}
\begin{aligned}
    A(\tau)=(1-\tau)Z \otimes I + \tau X\otimes A \\
    \ket{b'}= \frac{1}{\sqrt{2}}(\ket{0}+ \ket{1})\otimes \ket{b}. 
\end{aligned}
\end{equation}

\begin{algorithm}[!h]
    \caption{VQLS}
    \label{alg:AC2}
    \renewcommand{\algorithmicrequire}{\textbf{Input:}}
    \renewcommand{\algorithmicensure}{\textbf{Output:}}
    \begin{algorithmic}[1]
        \REQUIRE Coefficient matrix: $A$;
        Homogeneous term: $\vec{b}$;
        
        Initial parameters: $\vec{\theta}_0$; 
        Precision constraint: $\epsilon$; 
        
        Learning rate: $\beta$.  
        \ENSURE  Numerical solution $\vec{x}$   

	  \STATE Decompose $A$ into a linear combination of Pauli matrices;
	  \STATE Construct a quantum circuit of ansatz $U(\vec{\theta})$;
	   
 	\STATE Construct the quantum circuit set ${C_1(\vec{\theta})}$ corresponding to $\bra{b}A \ket{\psi}$ based on the ansatz $U(\vec{\theta})$ and the Pauli decomposition of $A$, along with $\vec{b}$;

	\STATE Construct the quantum circuit set ${C_2(\vec{\theta})}$ corresponding to $\frac{\partial \bra{\psi}}{\partial \theta_i} A^{\dagger}A \ket{\psi}$ based on the ansatz $U(\vec{\theta})$, ansatz derivatives, and the Pauli decomposition of $A$;

	\STATE Construct the quantum circuit set ${C_3(\vec{\theta})}$ corresponding to $\frac{\partial \bra{\psi}}{\partial \theta_i}A^{\dagger} \ket{b}$ based on the ansatz $U(\vec{\theta})$, the Pauli decomposition of $A$, and the vector $\vec{b}$;

	\STATE Set $\mathrm{criteria} = 1$;
        \WHILE{$\mathrm{criteria} > \epsilon$}
		\STATE Compute the current step's expected value $C(\vec{\theta})$ based on the quantum circuit sets ${C_0(\vec{\theta})}$ and ${C_1(\vec{\theta})}$;
		\STATE  Update $\mathrm{criteria} = C(\vec{\theta})$;
		\IF{$\mathrm{criteria} \leq \epsilon$} 
		\STATE break;
		\ENDIF
            \FOR{each $i \in [\vec{\theta}.\mathrm{size}]$}
			\STATE Compute $\nabla C_i(\vec{\theta}_t)$ based on the measurement results from the quantum circuit sets ${C_1(\vec{\theta})}$, $						{C_2(\vec{\theta})}$, and ${C_3(\vec{\theta})}$;
			\STATE Update $\theta^{t+1}_i = \theta^{t}_i - \beta \nabla C_i(\vec{\theta}_t) $;
		\ENDFOR
		\STATE Update the parameter information for the quantum circuit sets ${C_s(\vec{\theta})}$, where $s=0,1,2,3$, to $\vec{\theta}_{t+1}$;
		
        \ENDWHILE
		\STATE Obtain the optimal parameter information $\vec{\theta}_{\text{opt}}$;

		\STATE Get the quantum state information $\vec{x}_{\mathrm{norm}}$ of the ansatz $U(\vec{\theta})$ under the optimal parameter settings $\vec{\theta}_{\text{opt}}$ through quantum state tomography;

		\STATE Calculate the normalization factor $\eta;$

		\STATE Compute $\vec{x} = \eta \vec{x}_{\mathrm{norm}};$
	
        \RETURN  Numerical solution $\vec{x}.$
    \end{algorithmic}
\end{algorithm}

When the value of the cost function approaches 0, $\ket{\psi(\vec{\theta})}$ becomes a normalized solution in the quantum state. After extraction with the tomography method, the final solution $\vec{y}$ of the original linear equations can be recovered from the extracted normalized solution $\vec{y'}$ by 
\begin{equation}
\vec{y} = \eta \vec{y'}.
\label{eq:SUB-VQLS-1}
\end{equation}
To compute the coefficient $\eta$, substitute equation (\ref{eq:SUB-VQLS-1}) back to the linear equation system and yield
\begin{equation}
 \eta H \vec{y'} = \vec{r} .
\label{eq:SUB-VQLS-2}
\end{equation}
Next, both sides of the equation multiply their transposes respectively as follows
\begin{equation}
 \eta^2 \vec{y'}^T H^T H \vec{y'} = \vec{r}^T \vec{r} .
\label{eq:SUB-VQLS-3}
\end{equation}
The term $ \vec{y'}^T H^T H \vec{y'}$ on the left-hand side can be obtained by measuring the term $\bra{\psi} H^T H \ket{\psi}$ in the loss function. The term on the right-hand side equals $\Vert \vec{r} \Vert_2 $.

\section{Hybrid quantum-classical CFD framework} \label{pattern}

\subsection{Hybrid quantum-classical computing architecture}
Numerically solving Navier-Stokes equations involves a series of complex algorithms, including numerical schemes for spatial and temporal terms. Implicit methods are applied in time advancement and result in a large linear algebra system that is computationally expensive to solve. In the classical computing paradigm, these large linear systems can be efficiently solved with parallel techniques on CPU/GPU. In the hybrid quantum-classical computing framework, they are solved by the quantum linear solver on quantum computers. Similar to heterogeneous computing based on the current graphic processing unit~(GPU), the quantum processing unit~(QPU) can be regarded as an accelerator of classical computers. To realize collaborative quantum-classical computing for computational fluid dynamics, a hybrid quantum-classical architecture diagram is shown in Figure \ref{fig:cqhofcfd-architecture}. The hybrid quantum-classical computing architecture mainly consists of the following four layers:

\begin{figure*}[htbp]
	\centering								
	\resizebox{0.8\textwidth}{!}{\includegraphics{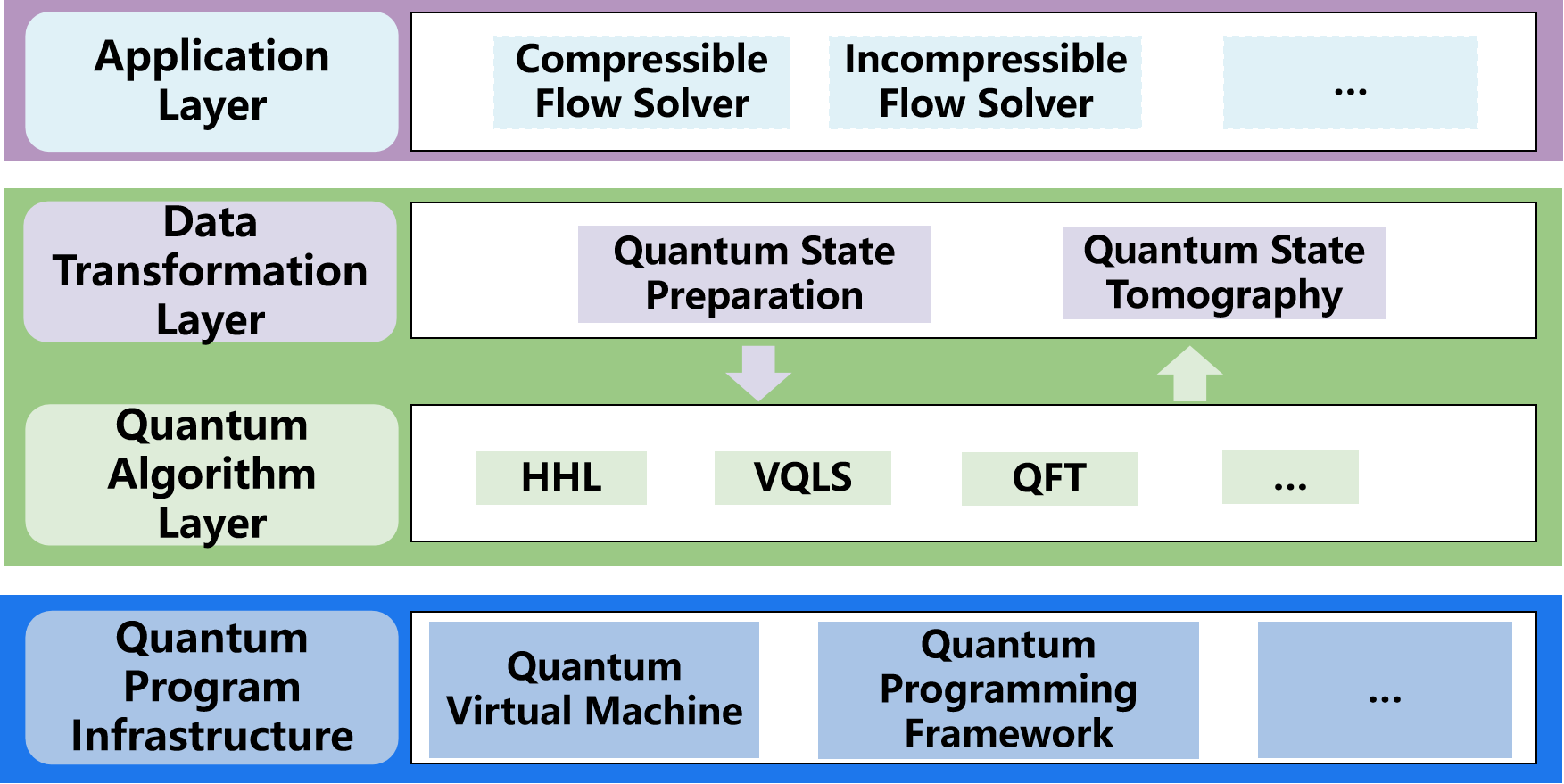}}	
	\caption{Hybrid quantum-classical computational fluid dynamics architecture diagram.}
    \label{fig:cqhofcfd-architecture}
\end{figure*}

\begin{itemize}
    \item \textbf{Quantum Program Infrastructure Layer} This is the bottom layer of the architecture and provides basic quantum programming infrastructures, including program framework, quantum simulators, etc. The quantum programming framework is supported by QPanda~\cite{dou2022qpanda}. Quantum simulators, called quantum virtual machines as well, include a built-in simulator in QPanda and a sparse state simulator integrated with quantum random accessing memory (QRAM) simulator~\cite{chen2023scalable}.
    \item \textbf{Quantum Algorithm Layer} In this layer, quantum linear algorithms are included, such as the  HHL and the VQLS. These algorithms can be used independently or combined with certain classical methods like the subspace methods. The algorithms are designed to be easily called by applications in various scientific computing fields.
    \item \textbf{Data Transformation Layer} Data conversions between quantum and classical algorithms are conducted in this layer, including quantum state preparation and quantum state tomography.
    \item \textbf{Application Layer} This is the top layer in classical computers. Various applications port computation tasks onto quantum computers by calling APIs provided by the quantum algorithm layer.
\end{itemize}

\begin{figure*}[htbp]
    \centering								
    \resizebox{0.618\textwidth}{!}{\includegraphics{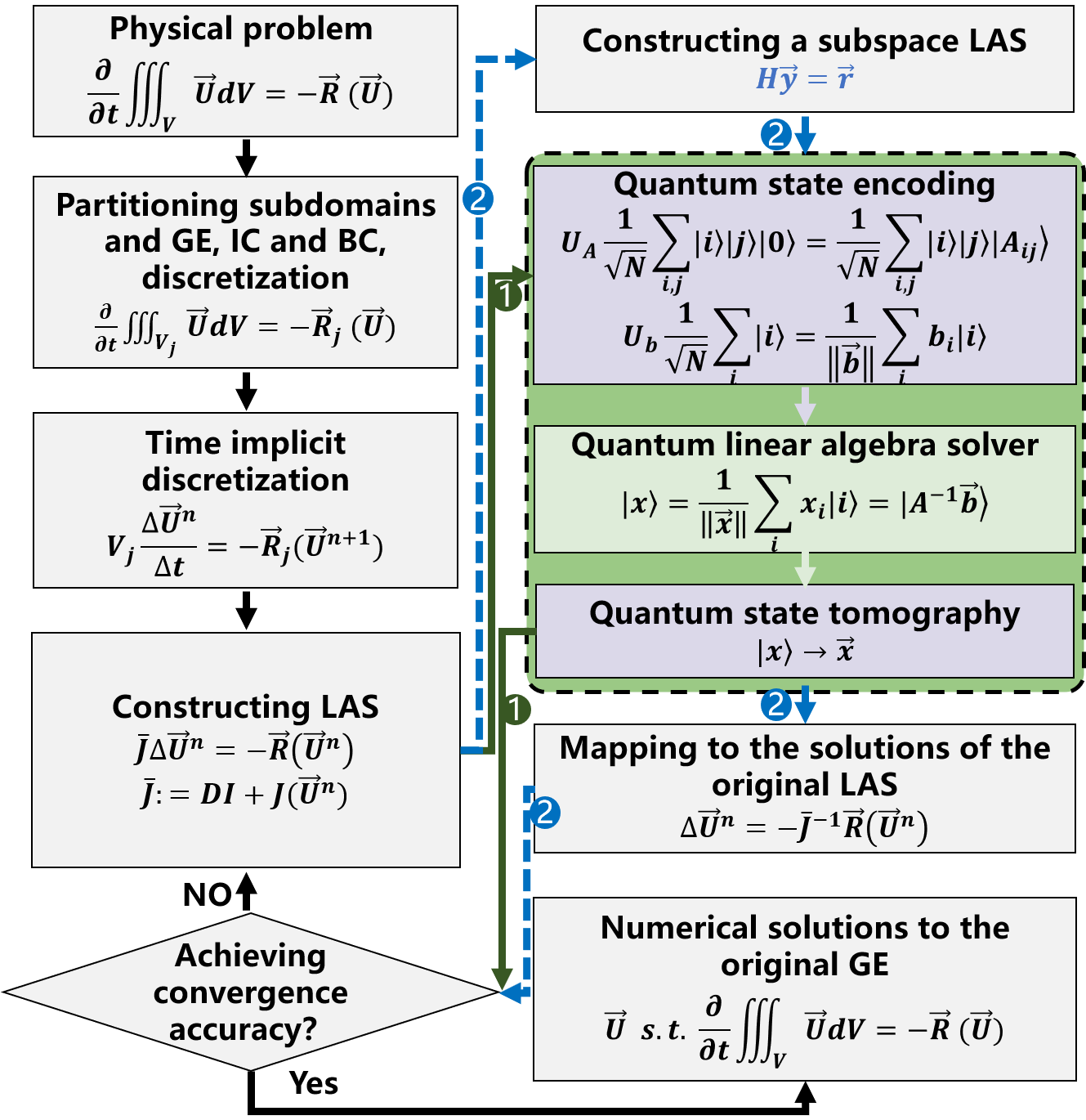}}	
    \caption{Flowchart of a hybrid quantum-classical CFD architecture. The processes in the green box are the core of the quantum linear solver and run on quantum devices. Other processes outside the green box are executed on classical computing devices. The solid lines marked as "1" solve the original linear system directly with a quantum linear solver, while the dashed lines marked as "2" are a quantum linear solver combined with classical subspace methods. (GE, IC, BC, and LAS stand for governing equations, initial conditions, boundary conditions, and linear algebraic systems, respectively)}
    \label{qcfd-flow-diagram}
\end{figure*}

Figure \ref{qcfd-flow-diagram} shows the flowchart of the proposed hybrid quantum-classical CFD architecture. The left-hand side of the flowchart contains the main process of CFD, while the right-hand side is the main body of the quantum linear solver (QLS). 

There are two ways to solve a linear system. The first approach, marked by a solid green line, uses standard quantum linear solvers like the HHL and VQLS to solve the original linear system directly. The matrix and vector of the linear system are encoded into quantum states with specific encoding methods. When the quantum linear solver is complete, the solution is read out through quantum state tomography. The second approach is marked with a dashed blue line. In fact, this is an improved QLS proposed in our previous work~\cite{chen2024enabling} and is called SUB-QLS. In SUB-QLS, the standard QLS is embedded into the Krylov subspace method to solve a low-dimensional linear system. More details of SUB-QLS will be given in the next section. Due to the nonlinearity of the discretized equations, a Newton iteration procedure is needed to find the true solution, thus the solution of the linear system found above needs to satisfy the convergence criteria of the Newton iteration as well.

\subsection{QLAS library}
The quantum linear algebra solver (QLAS) is the key component of the current hybrid quantum-classical CFD architecture. It covers two middle layers in Figure \ref{fig:cqhofcfd-architecture}, including the quantum encoding of the matrix and the right-hand side vector into quantum states, performing quantum linear algorithms, and reading out quantum states. The QLAS library is designed in the quantum programming framework QPanda and is based on quantum logic gates. Solvers inside can run on both real quantum devices and simulators.

\begin{figure*}[htbp]
  \centering
  \includegraphics[width=0.75\textwidth]{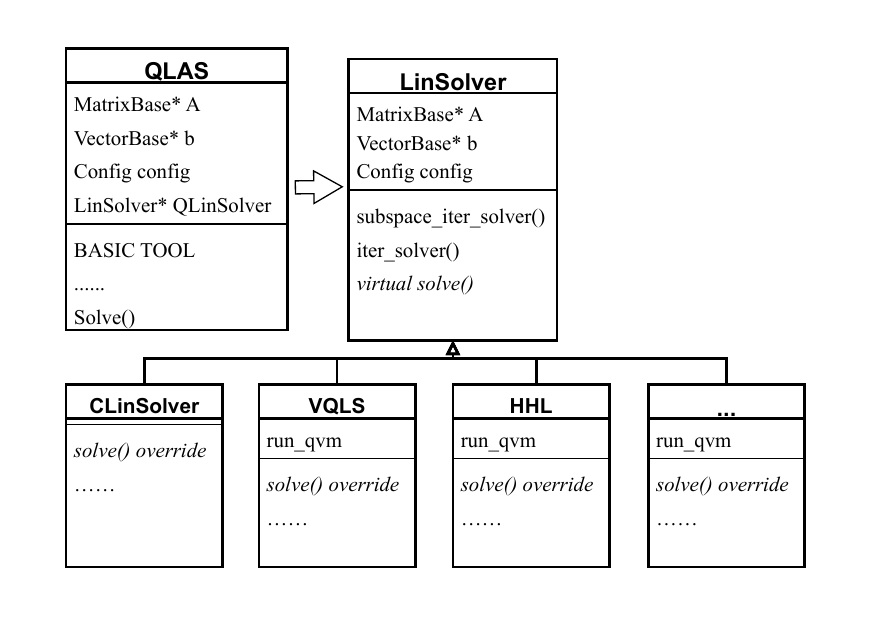}
  \caption{Class diagram of the QLAS library.}
  \label{FigQLAS-class}
\end{figure*}

Figure \ref{FigQLAS-class} depicts the class diagram of the QLAS library. The core module is the abstract class \texttt{LinSolver} and its related derived classes. The abstract class \texttt{LinSolver} includes member variables and member functions that are responsible for the quantum-classical solving of linear systems. The virtual function \texttt{solve()} executes the solving process, which is overridden by subclasses of difference linear algorithms, including classical methods like CG and quantum methods like HHL and VQLS. The function \texttt{iter\_solver()} is an iterative solver, while \texttt{subspace\_iter\_solver()} is a subspace iterative solver. Both solvers are associated with the solutions returned by the function \texttt{solve()}.  Data in QLAS is managed by the \texttt{MatrixBase} class and the \texttt{VectorBase} class which support many popular dense and sparse storage methods. 

The QLAS library is designed based on the strategy design pattern to promote code reusability, flexibility, and maintainability. 

\subsection{Subspace quantum linear solver}
It is still challenging for both types of quantum linear algorithms to perform large-scale fluid simulations on existing quantum computing devices, including real quantum computers and quantum simulators, because the number of quantum logic gates increases dramatically with problem size. To scale the problem size on current quantum devices, it is necessary to reduce the dimension of the linear system solved with quantum linear solvers. Subspace methods offer paths to do so. Recently, quantum Krylov subspace method~\cite{stair2020multireference,jain2021krylov,cortes2022krylov} was introduced into the HHL algorithm and presented as an alternative to quantum phase estimation (QPE) for eigenpair problems, in which deep quantum circuits were replaced with multiple shallow ones~\cite{xu2024quantum}. This method has been numerically tested on systems with dimensions up to $2^{10} \times 2^{10}$. Another application of the subspace method was SUB-QLS~\cite{chen2024enabling}. The core idea is to combine a standard QLS with the Krylov subspace method and find the solution in the subspace. The standard QLS is called repeatably with an iterative restart mechanism until it is converged. With the subspace method, a high-dimensional linear system can be transformed into a smaller one that is easier to solve with fewer quantum resources, thus a problem can be scaled to large sizes. This method has been successfully applied to solve a 5043-dimensional linear system and performed the largest fluid simulation on a near-term superconducting quantum computer~\cite{chen2024enabling}. The subspace method is performed on classical computers here, but it can be performed on quantum computers as well~\cite{jain2021krylov}. 

Algorithm \ref{alg:AC3} shows how the standard quantum HHL and the VQLS algorithms are integrated with the Krylov subspace method to construct the SUB-QLS solvers, which will be called SUB\_HHL and SUB\_VQLS.

\begin{algorithm}[!h]
    \caption{SUB\_HHL(VQLS) Algorithm}
    \label{alg:AC3}
    \renewcommand{\algorithmicrequire}{\textbf{Input:}}
    \renewcommand{\algorithmicensure}{\textbf{Output:}}
    \begin{algorithmic}[1]
        \REQUIRE The matrix $A$, the vector $\vec{b}$, initial value$\vec{x}_0$, convergence precision $\epsilon$, Maximum subspace dimension $k$
	  \ENSURE  Numerical solution $\vec{x}$
        
	   \STATE Compute $\vec{r}_0 = \vec{b} - A \vec{x}_0$ and $\beta = \Vert \vec{r}_0 \Vert$
	   \STATE $\vec{v}_1 = \vec{r}_0 / \beta$, $\vec{r} = \beta \vec{e}_1$
         \STATE  $j = 1$
        \FOR{each $j \in [1,k]$}
		\IF {$\beta < \epsilon \sqrt{\Vert \vec{b} \Vert}$} 
		\STATE break
		\ENDIF
		\STATE $\vec{w}_j = A \vec{v}_j$
		\FOR{each $i \in [1,j]$}
		\STATE $h_{ij} = (\vec{w}_j, \vec{v}_i)$; $\vec{w}_j = \vec{w}_j - h_{ij}\vec{v}_i$
		\ENDFOR
		\STATE $h_{j+1,j} = \Vert \vec{w}_{j} \Vert$; $\vec{v}_{j+1} = \vec{w_j} / h_{j+1,j}$
		\FOR {each $i \in [1,j-1]$}
		\STATE $t = c_i h_{i,j} + s_i h_{i+1,j}$; $h_{i+1,j} = s_i h_{i,j} - c_i h_{i+1, j}$; $h_{i,j} = t$		
		\ENDFOR
           \IF {$|h_{j+1,j}| = 0$}
			\STATE $c_j = 1 $, $s_j = 0$
		\ELSE
			\STATE $t = h_{j+1,j} / h_{j,j}$
			\IF {$|t| > 1$}
				\STATE $c_j = 1/\sqrt{1+t^2}$; $s_j = c_j t$
			\ELSE
				
            		\STATE $t = h_{j,j} / h_{j+1,j}$;$s_j = 1/ \sqrt{1+t^2}$, $c_j = s_j t$
				
			\ENDIF
            \ENDIF
		\STATE $t = c_j r_j, r_{j+1} = s_j r_{j}; r_j = t$	
  \STATE $h_{j,j} = c_j h_{j,j} + s_j h_{j+1,j}, h_{j+1,j} = 0$
          \STATE $\beta = \left| r_{j+1} \right|$
	
        \ENDFOR
		 \STATE Solve the new linear subspace system($H_j \vec{y} = \vec{r}(1:j)$) using the HHL (VQLS) linear algorithm 
   
		\STATE Update $\vec{x} = \vec{x}_0 + V_j \vec{y}$,where $V_j = [\vec{v}_1, \cdots , \vec{v}_j]$
        
        \RETURN The numerical solution is $\vec{x}$.
    \end{algorithmic}
\end{algorithm}

\section{Numerical cases} \label{NumExp}
In this section, several test cases are performed to validate the feasibility of the hybrid quantum-classical CFD framework and the correctness of quantum linear algorithms in CFD. The popular open-source CFD solver SU2 is chosen as the test CFD solver. Limited to the classical computing resource, quantum linear solvers integrated with the subspace method are adopted in all cases. For convenience, the HHL and VQLS algorithms with the subspace method are labeled as SUB\_HHL and SUB\_VQLS respectively. Meanwhile, for all cases, solutions computed with built-in classical linear solvers in SU2 are used to be reference solutions to validate the quantum solutions.

\subsection{Inviscid Bump}
In this case, a flow passing an inviscid bump in a channel is simulated~\cite{SU2-Inviscid-Bump}. The flow is governed by the two-dimensional compressible Euler equations. The Computational domain is divided into 32385 non-overlapped grid cells. For the numerical method, the second-order JST scheme is used in space discretization, and the temporal terms are discretized with the first-order implicit Euler scheme. Moreover, the incomplete LU (ILU) preconditioner is adopted to lower the condition number of the linear algebra system, and the dimension of the subspace is set to 8. All cases stopped when the same convergence criteria were satisfied.




\begin{figure*}[tbh]
    \centering
    \subfigure {\
        \begin{minipage}[b]{0.42\textwidth}
            \centering
                \begin{overpic}[scale=0.35]{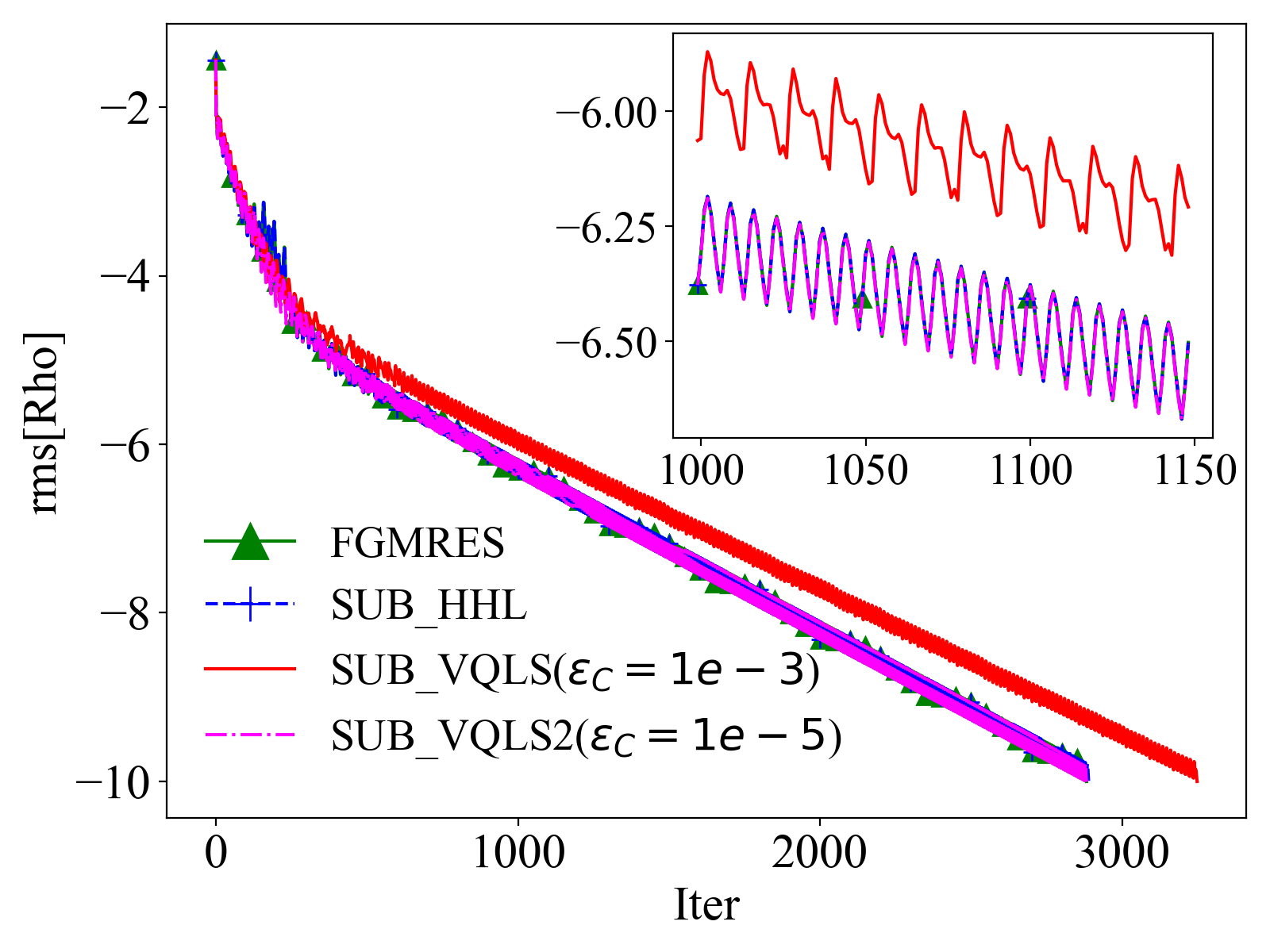}
                \put(3,72){(a)}
            \end{overpic}
        \end{minipage}
    }
    \subfigure {\
    \begin{minipage}[b]{0.42\textwidth}
            \centering
               \begin{overpic}[scale=0.35]{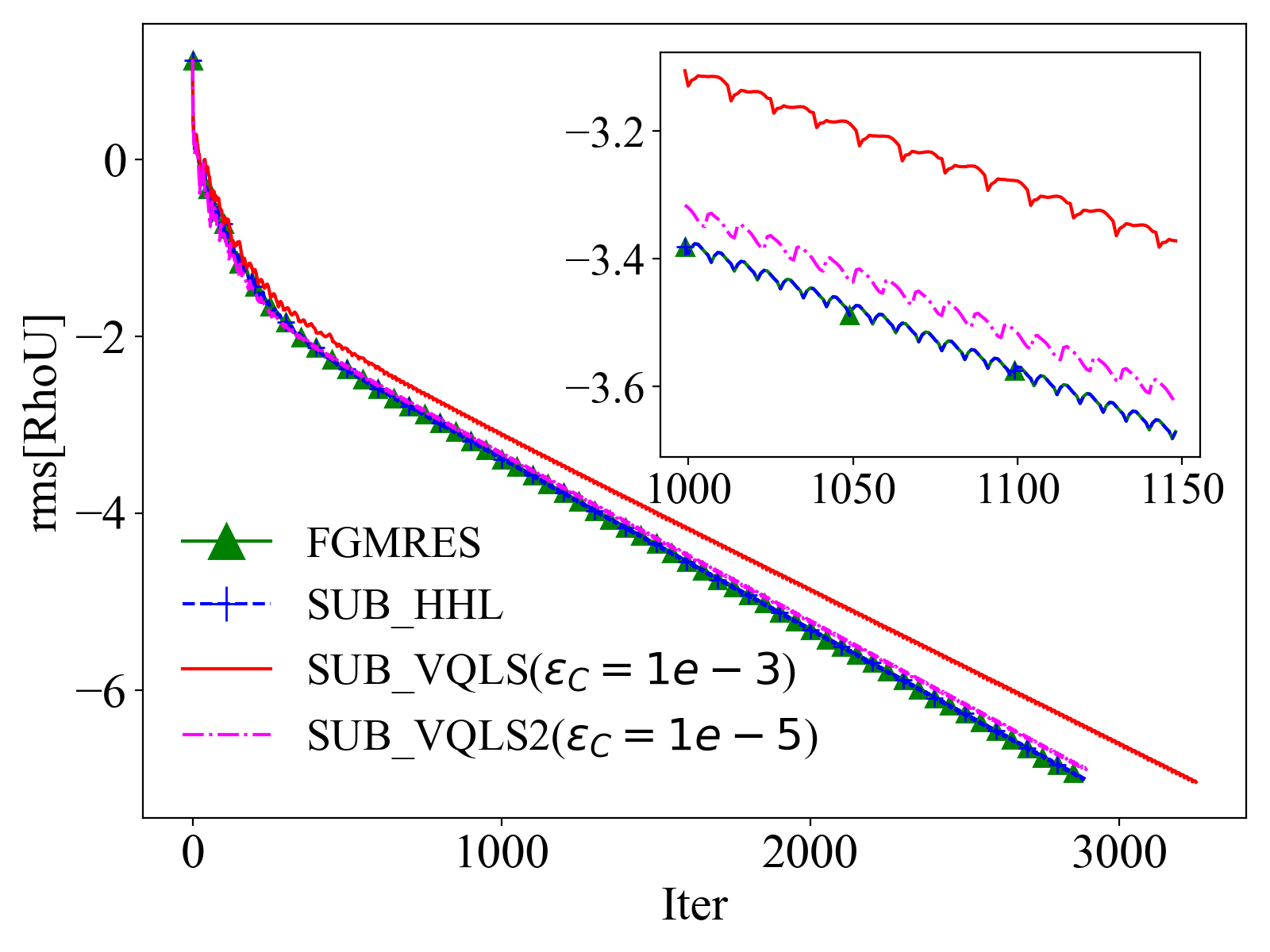}
               \put(3,72){(b)}
            \end{overpic}
        \end{minipage}
    }
        \\
    \subfigure {\
    \begin{minipage}[b]{0.42\textwidth}
            \centering
               \begin{overpic}[scale=0.35]{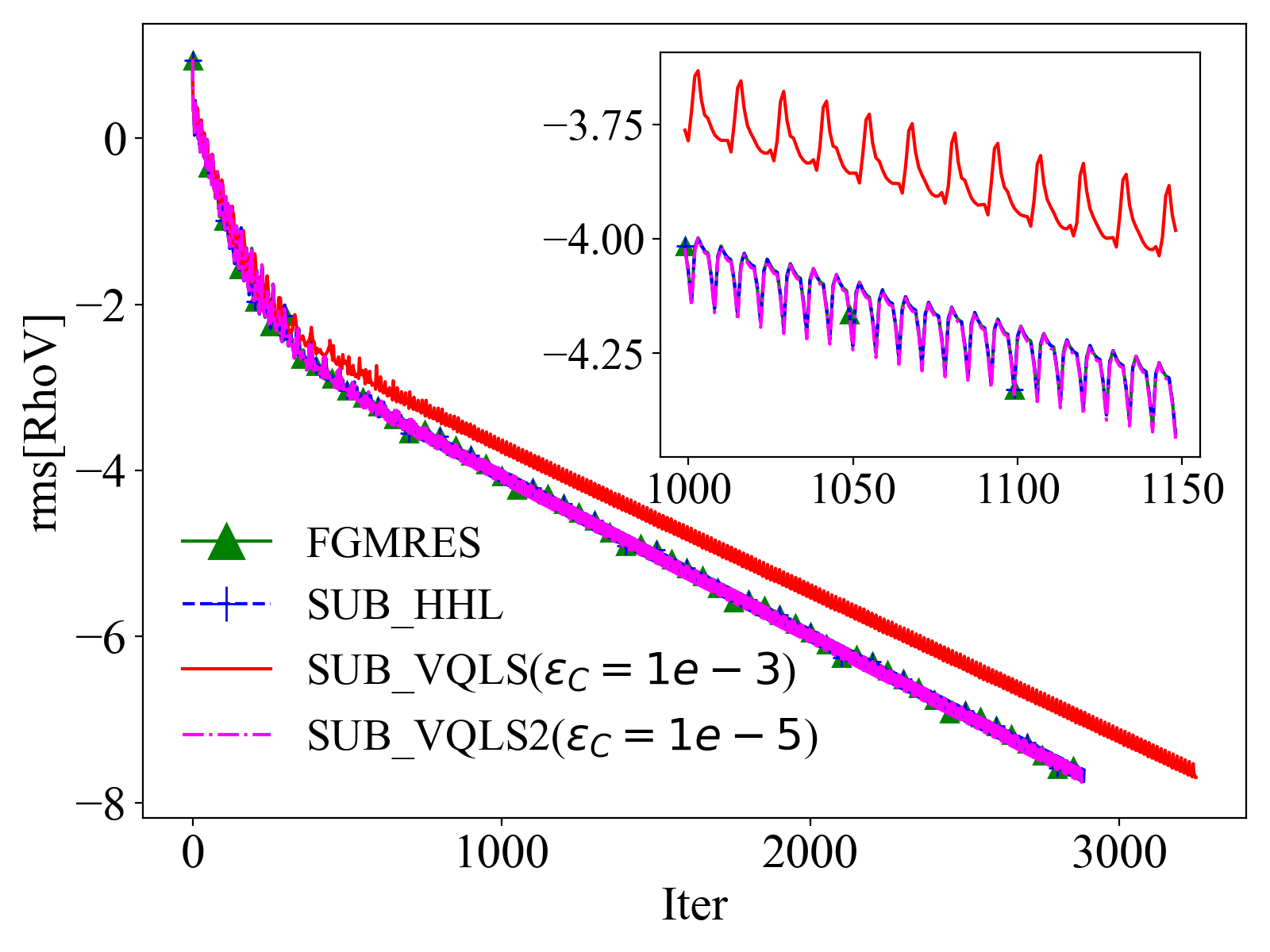}
               \put(3,72){(c)}
            \end{overpic}
        \end{minipage}
    }     
    \subfigure {\
    \begin{minipage}[b]{0.42\textwidth}
            \centering
               \begin{overpic}[scale=0.35]{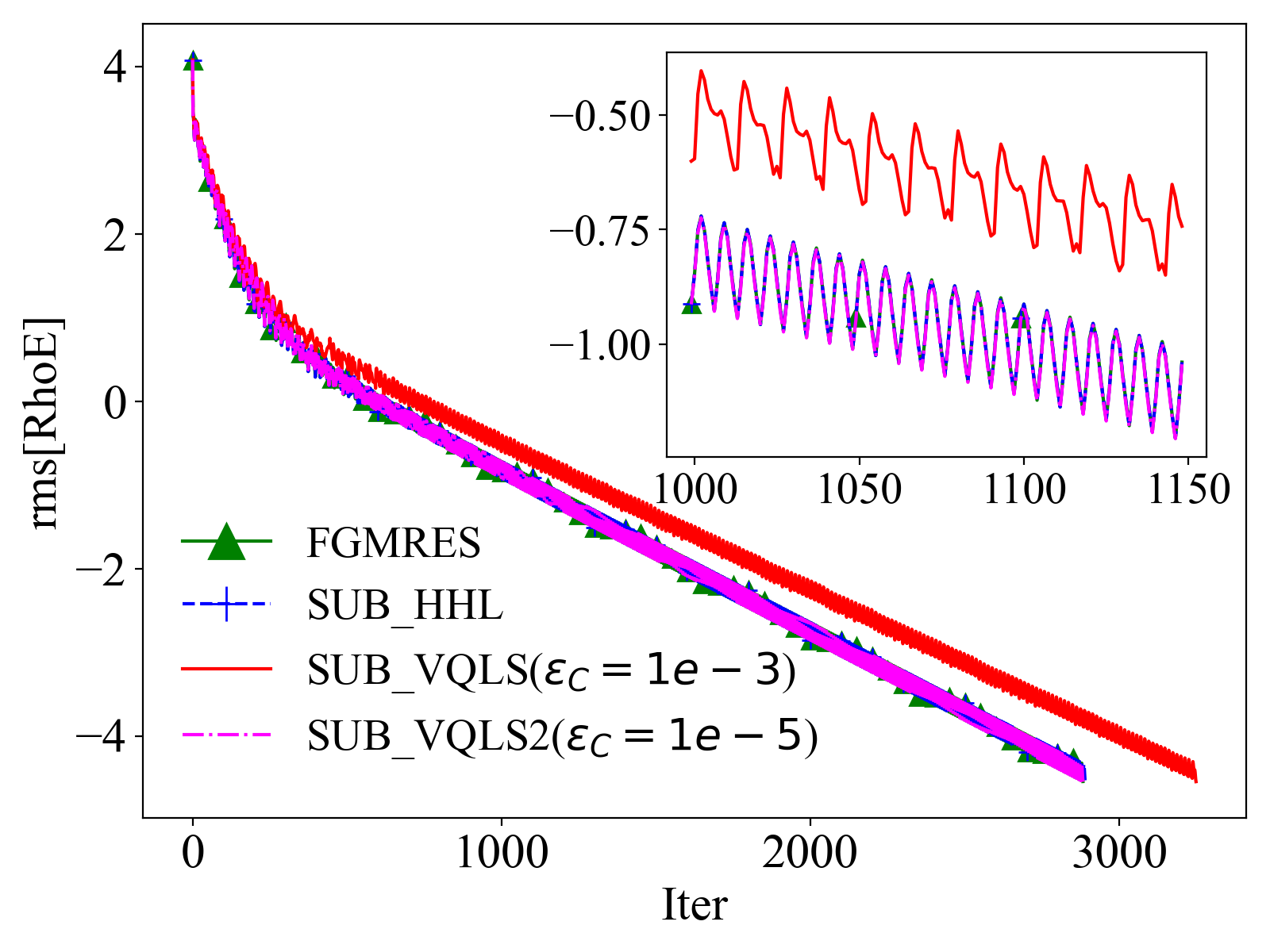}
               \put(3,72){(d)}
            \end{overpic}
        \end{minipage}
    }
    \caption{Convergence histories of residuals. The green solid lines with triangles represent data of the classical FGMRES method, while the blue dashed lines with cross markers for SUB\_HHL, and the red solid lines for SUB\_VQLS.}
    \label{fig:Inviscid-Bump-rms}
\end{figure*}

Because the VQLS algorithm is an iterative method, convergence precision is needed to stop the iteration. In this case, convergence precision $\epsilon_C = 10^{-3}$ and $10^{-5}$ are chosen to investigate the effect of precision on the convergence of the CFD solver. The convergence history is shown in Figure \ref{fig:Inviscid-Bump-rms}. It is shown that the classical method and the SUB\_HHL method have nearly the same convergence characteristic because they are direct methods for solving linear equations, so they get very accurate results if the linear system is non-singular. For the SUB\_VQLS method, when the convergence precision is high enough, it behaves like the direct methods. In contrast, lower convergence precision leads to slower convergence of the CFD solver and needs more outer iterations. For quantum resource requirements, the HHL method needs 19 qubits on average, while the VQLS method needs only 4 qubits. 

\begin{figure}[tbph]
    \centering
    \subfigure {\
        \begin{minipage}[b]{0.42\textwidth}
            \centering
                \begin{overpic}[scale=0.35]{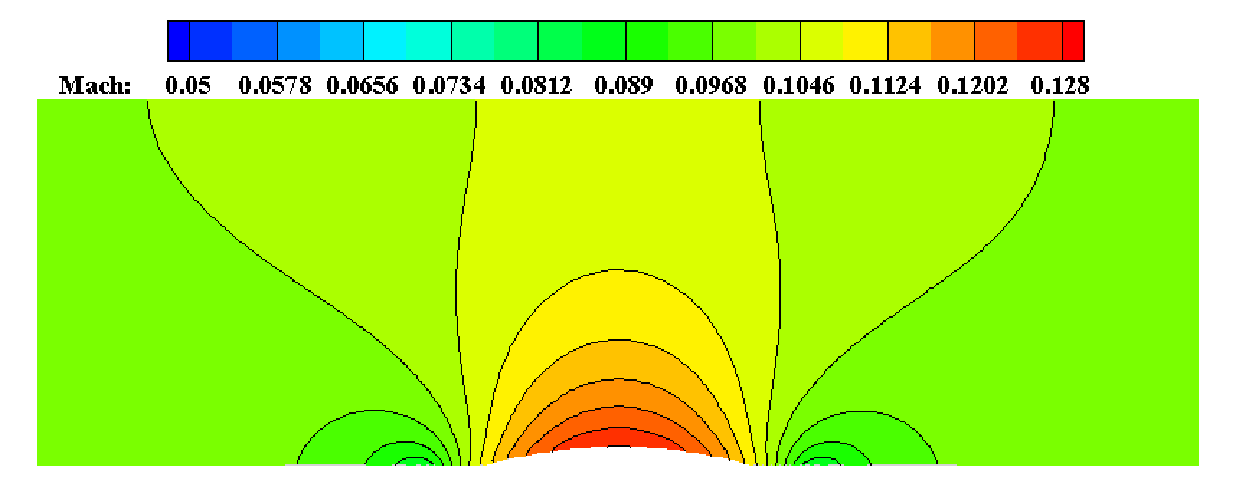}
                \put(3,38){(a)}
                \put(5,26){\fontsize{8pt}{12pt}\selectfont SUB\_HHL}
            \end{overpic}
        \end{minipage}
    }
    \\
    \subfigure {\
    \begin{minipage}[b]{0.42\textwidth}
            \centering
               \begin{overpic}[scale=0.35]{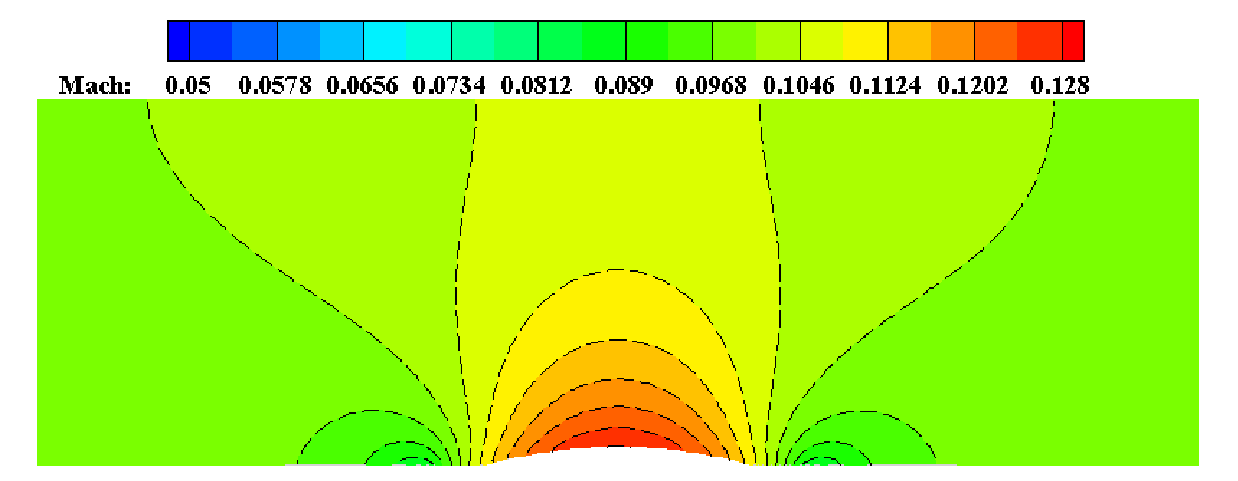}
               \put(3,38){(b)}
               \put(5,26){\fontsize{8pt}{12pt}\selectfont SUB\_VQLS}
            \end{overpic}
        \end{minipage}
    }
    \caption{ The Mach number distribution comparison for classical and quantum solvers(SUB\_HHL and SUB\_VQLS).The flood distribution is the result of classical solver FGMRES, and the contour lines represent the quantum results. (a) Comparison result of classical FGMRES method and the quantum SUB\_HHL method.  (b) Comparison result of classical FGMRES method and the quantum SUB\_VQLS method. }
    \label{FigInviscid-Bum-Mach}
\end{figure}

\begin{figure}[tbph]
    \centering
    \subfigure {\
        \begin{minipage}[b]{0.42\textwidth}
            \centering
            \begin{overpic}[scale=0.35]{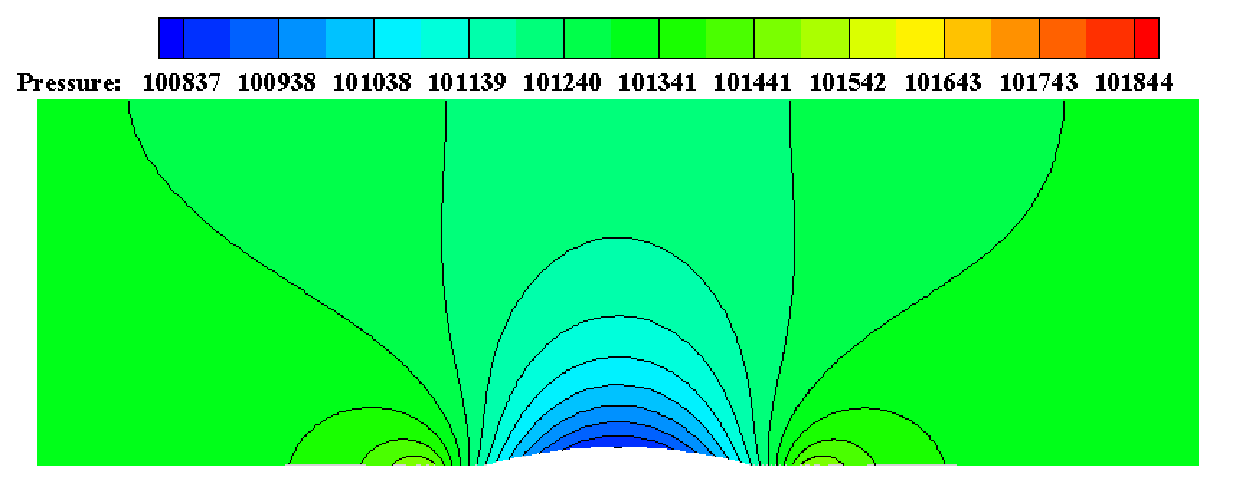}
                \put(3,38){(a)}
                \put(5,26){\fontsize{8pt}{12pt}\selectfont SUB\_HHL}
            \end{overpic}
        \end{minipage}
    }
    \\
    \subfigure {\
    \begin{minipage}[b]{0.42\textwidth}
            \centering
               \begin{overpic}[scale=0.35]{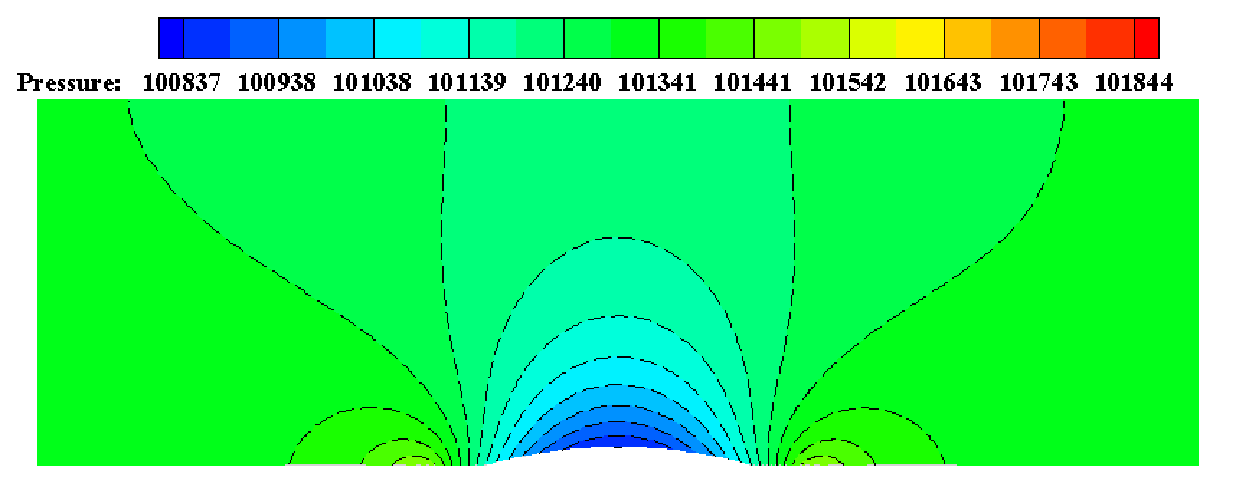}
               \put(3,38){(b)}
               \put(5,26){\fontsize{8pt}{12pt}\selectfont SUB\_VQLS}
            \end{overpic}
        \end{minipage}
    }
    \caption{ The pressure distribution comparison for classical and quantum solvers(SUB\_HHL and SUB\_VQLS). The flood distribution is the result of classical solver FGMRES, and the contour lines represent the quantum results. (a) Comparison result of classical FGMRES method and the quantum SUB\_HHL method.  (b) Comparison result of classical FGMRES method and the quantum SUB\_VQLS method. }
    \label{FigInviscid-Bum-pressure}
\end{figure}

\begin{figure}[ht]
	\centering								
	\includegraphics[scale=0.4]{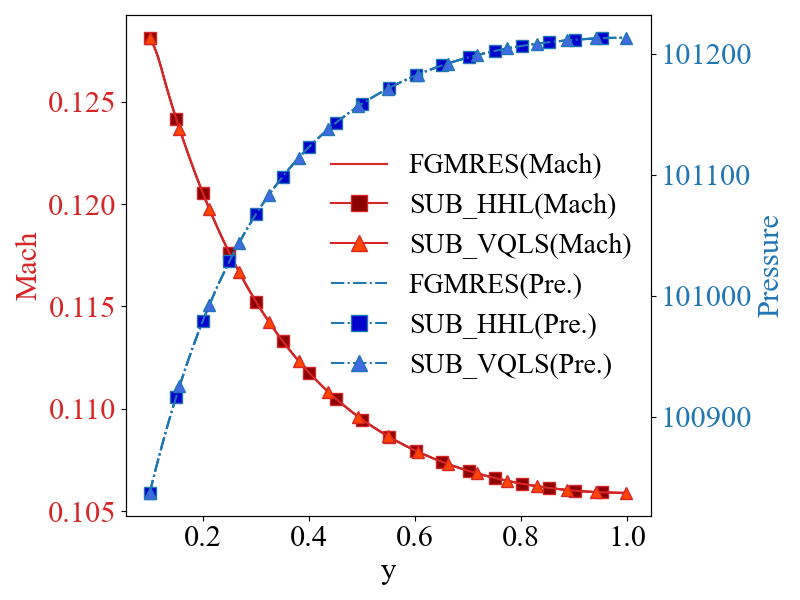}		
	\caption{The Mach number and pressure distribution comparison of the inviscid bump at cross-section x = 1.5.}
	\label{fig:inv_bump_x=1.5}
\end{figure}

Figure \ref{FigInviscid-Bum-Mach} and Figure \ref{FigInviscid-Bum-pressure} show the Mach number and pressure distribution contour comparison between the classical linear solver and the quantum linear solvers. Moreover, distribution on a cross-section at $x=1.5$ is shown in Figure \ref{fig:inv_bump_x=1.5} for better quantitative comparisons. All results of quantum linear solvers agree pretty well with those of classical solvers, which indicates that the quantum linear algorithms are feasible in CFD simulation. 

\subsection{Unsteady Laminar Cylinder Flow}
\begin{figure}[htbp]
	\centering								
	\includegraphics[scale=0.33]{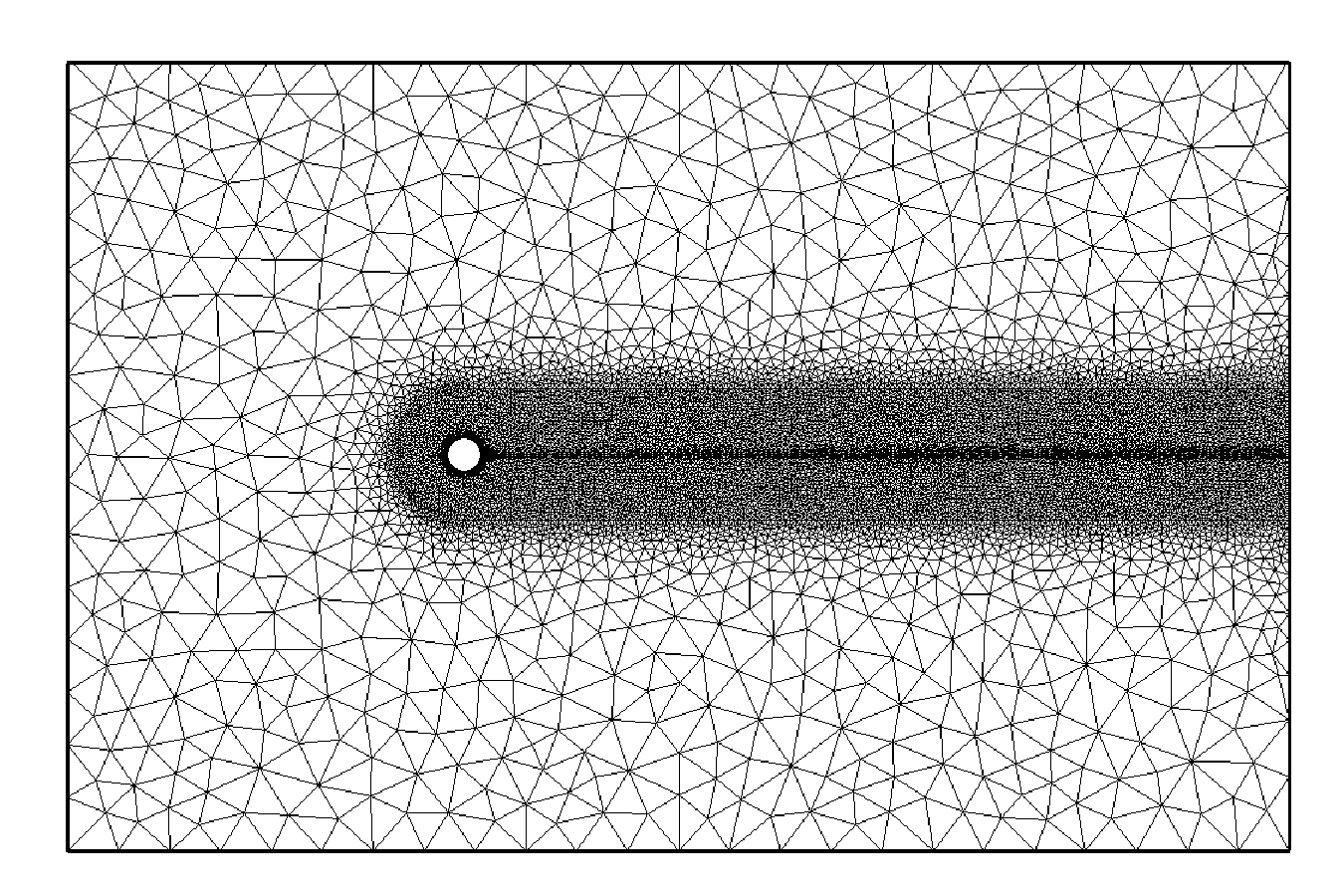}		
	\caption{Computational domain of circular cylinder flow.}
	\label{fig:grid_cylinder}
\end{figure}

A laminar flow past a circular cylinder is one of the most typical cases to validate the temporal scheme. When the Reynolds number is larger than the critical value, the famous Karman vortex street with a specific frequency is generated in the wake. In this case, a Mach 0.3 flow passes a cylinder, and the Reynolds number based on the states of the incoming flow and the diameter of the cylinder is 1000. The diameter of the circular cylinder is 1.0m. The two-dimensional Navier-Stokes equations are solved in the computational domain as depicted in Figure \ref{fig:grid_cylinder}. To capture the shedding vortex better, the mesh in the wake is refined, resulting in nearly 23 thousand grid cells.

The convective terms are computed with the Roe scheme, and the time is advanced with the second-order dual-time stepping method. The dimension of the subspace method is set to 8 for all classical and quantum cases. For all cases, the physical time step size is 0.0005s. 

\begin{figure}[h]
    \centering
    \subfigure[Time history of $C_d$. ]{\includegraphics[width=0.47\textwidth]{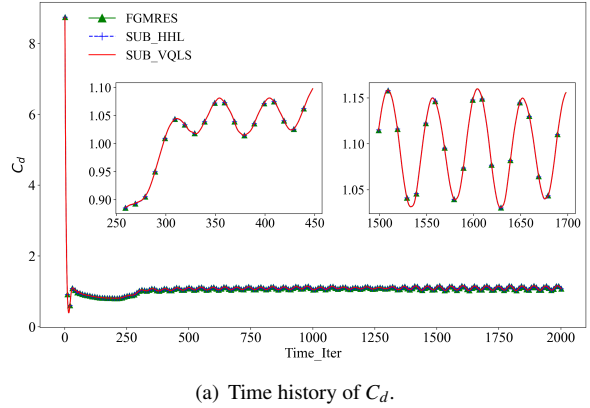}}
    \quad
    \subfigure[Time history of $C_l$.]{\includegraphics[width=0.47\textwidth]{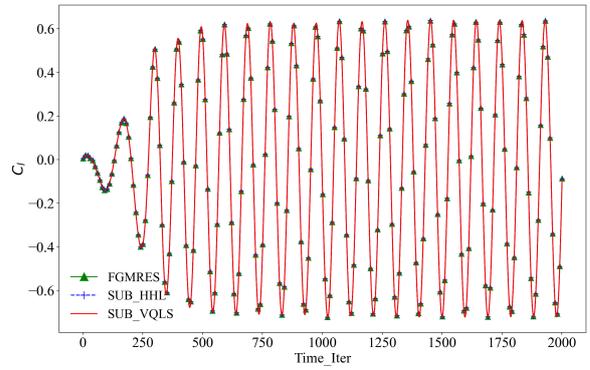}}
    \caption{Time histories of drag and lift coefficients.}
    \label{fig:cd_cl}
\end{figure}

The time histories of drag and lift coefficients are shown in Figure \ref{fig:cd_cl}. Firstly, we can get the period of the vortex shedding from the data after the flow is fully developed into a periodical state. The period is around 0.0475s. According to the definition of non-dimensional frequency, that is Strouhal number ($St=\frac{f \times D}{U} \approx 0.205$), which is consistent with the fact that the Strouhal number of flows past a circular cylinder varies between 0.18 and 0.22 over Reynolds number from 100 to 100000. Moreover, it can be seen that, for drag and lift coefficients, the quantum results coincide with the classical results during time advance, which proves the reliability of quantum algorithms in unsteady simulations.  

\begin{figure*}
    \centering
    \subfigure {\
        \begin{minipage}[b]{0.45\textwidth}
            \centering
            \begin{overpic}[scale=0.35]{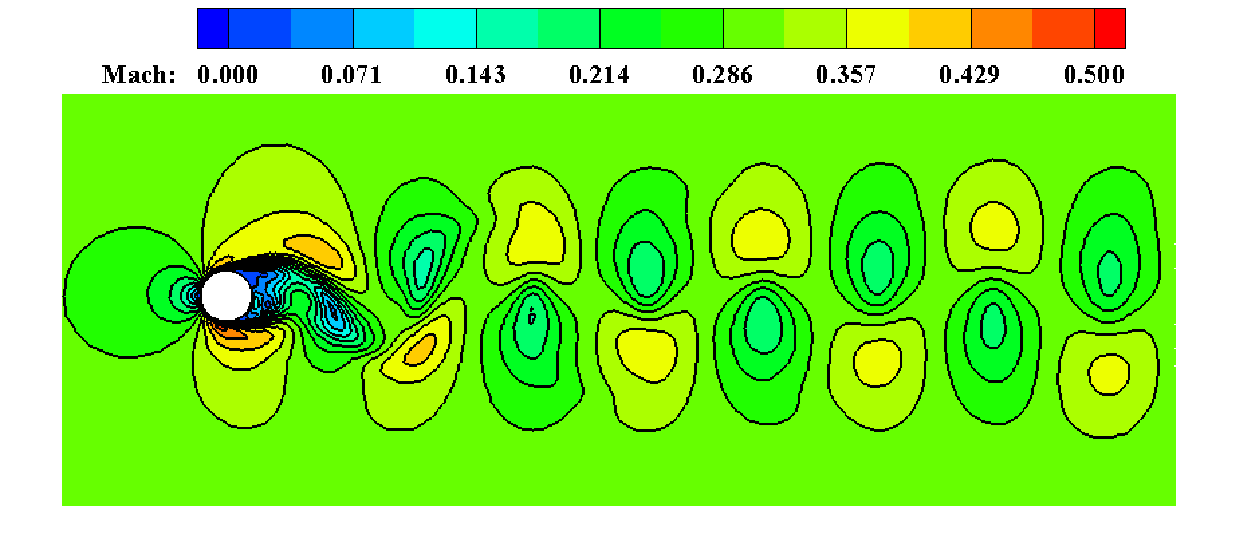}
                \put(3,42){(a)}
                \put(7,32){\fontsize{8pt}{12pt}\selectfont SUB\_HHL}
            \end{overpic}
        \end{minipage}
    }
    \subfigure {\
        \begin{minipage}[b]{0.45\textwidth}
            \centering
            \begin{overpic}[scale=0.35]{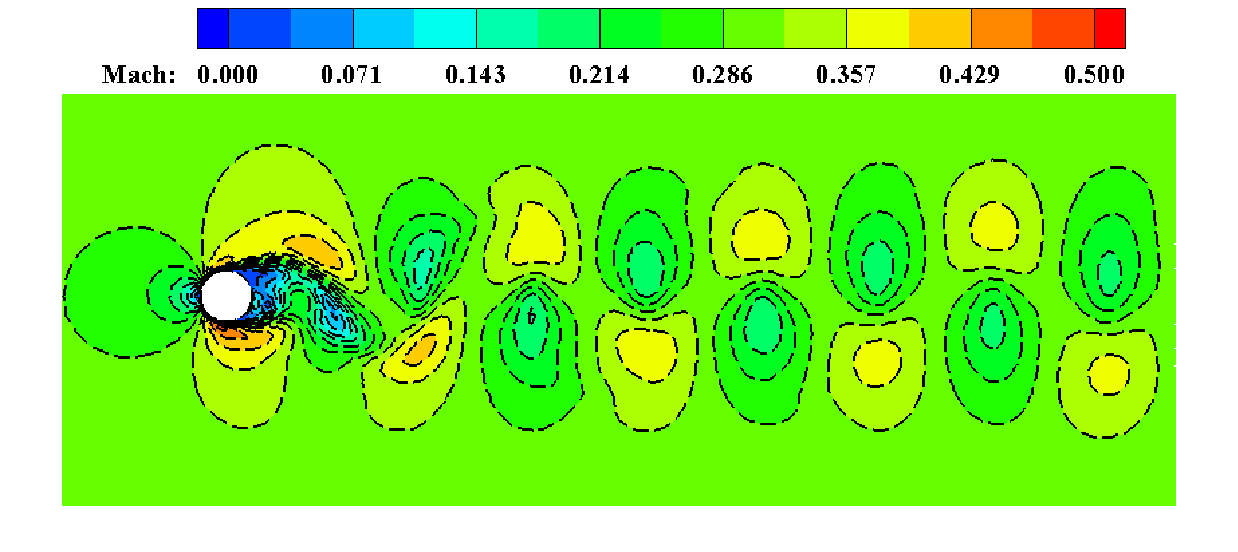}
                \put(3,42){(b)}
                \put(7,32){\fontsize{8pt}{12pt}\selectfont SUB\_VQLS}
            \end{overpic}
    \end{minipage}
    }
    \caption{ The Mach distribution comparison (t= 0.75s) for classical and quantum solvers (SUB\_HHL and SUB\_VQLS). The flood distribution is the result of classical solver FGMRES, and the contour lines represent the quantum results. (a) Comparison result of classical FGMRES method and the quantum SUB\_HHL method.  (b) Comparison result of classical FGMRES method and the quantum SUB\_VQLS method. }
    \label{FigUnsteadyCylinder-Mach}
\end{figure*}

Figure \ref{FigUnsteadyCylinder-Mach} shows a flow field snapshot's Mach number distribution comparison between the classical and quantum methods. Good agreement is observed, which proves the reliability of the quantum algorithms in unsteady CFD simulation.

\subsection{CHN-F1 Aircraft}
The CHN-F1 aircraft model is one of the publicly released models of the National Space Science and Data Center of China~\cite{CHN-F1}, and there is a lot of experiment data in different conditions that can be used to validate CFD solvers. Here, the  CHN-F1 model is used to validate the feasibility of the proposal framework and the quantum algorithms in practical engineering cases. The model is put in a Mach 0.6 flow, and the Reynolds number is 6266947.5 based on the reference length 0.5032m. The freestream temperature is 300K. The angle of attack (AoA) is ranged from 0 to 12 degrees.

Due to the high Reynolds number, the Splart-Allmaras turbulence model is applied to reduce the computation expanse.  Half of the model is simulated because of the geometric symmetry. The three-dimensional computational domain is divided into about 9.4 million grid cells. 

The JST scheme is adopted for convective terms, and the first-order Euler implicit scheme is used for temporal discretization. The dimension of subspace in solving a linear system is set to be 8.  

\begin{figure*}
    \centering
    \subfigure {\
        \begin{minipage}[b]{0.45\textwidth}
            \centering
                \begin{overpic}[scale=0.315]{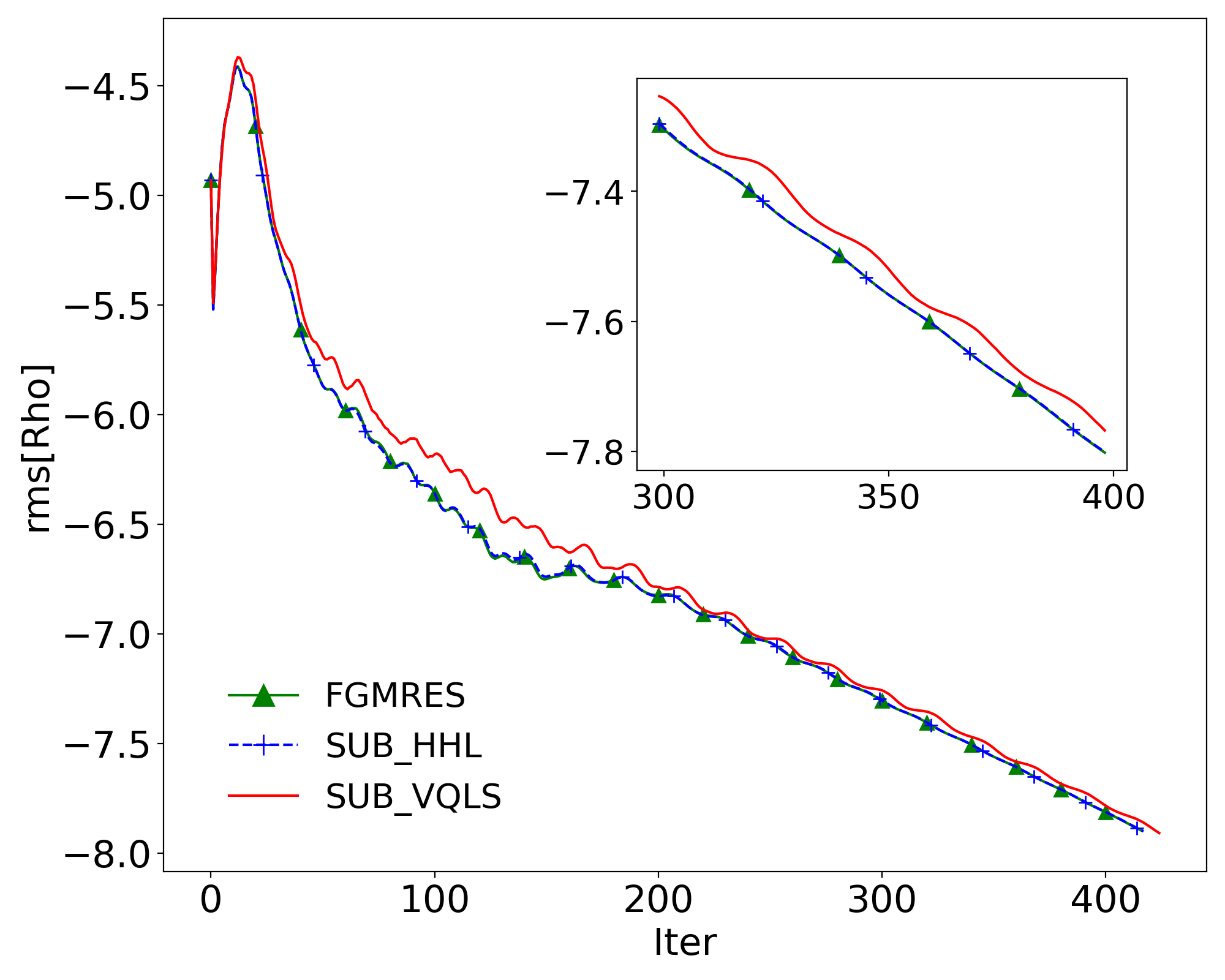}
                \put(3,76){(a)}
            \end{overpic}
        \end{minipage}
    }
    \subfigure {\
    \begin{minipage}[b]{0.45\textwidth}
            \centering
               \begin{overpic}[scale=0.315]{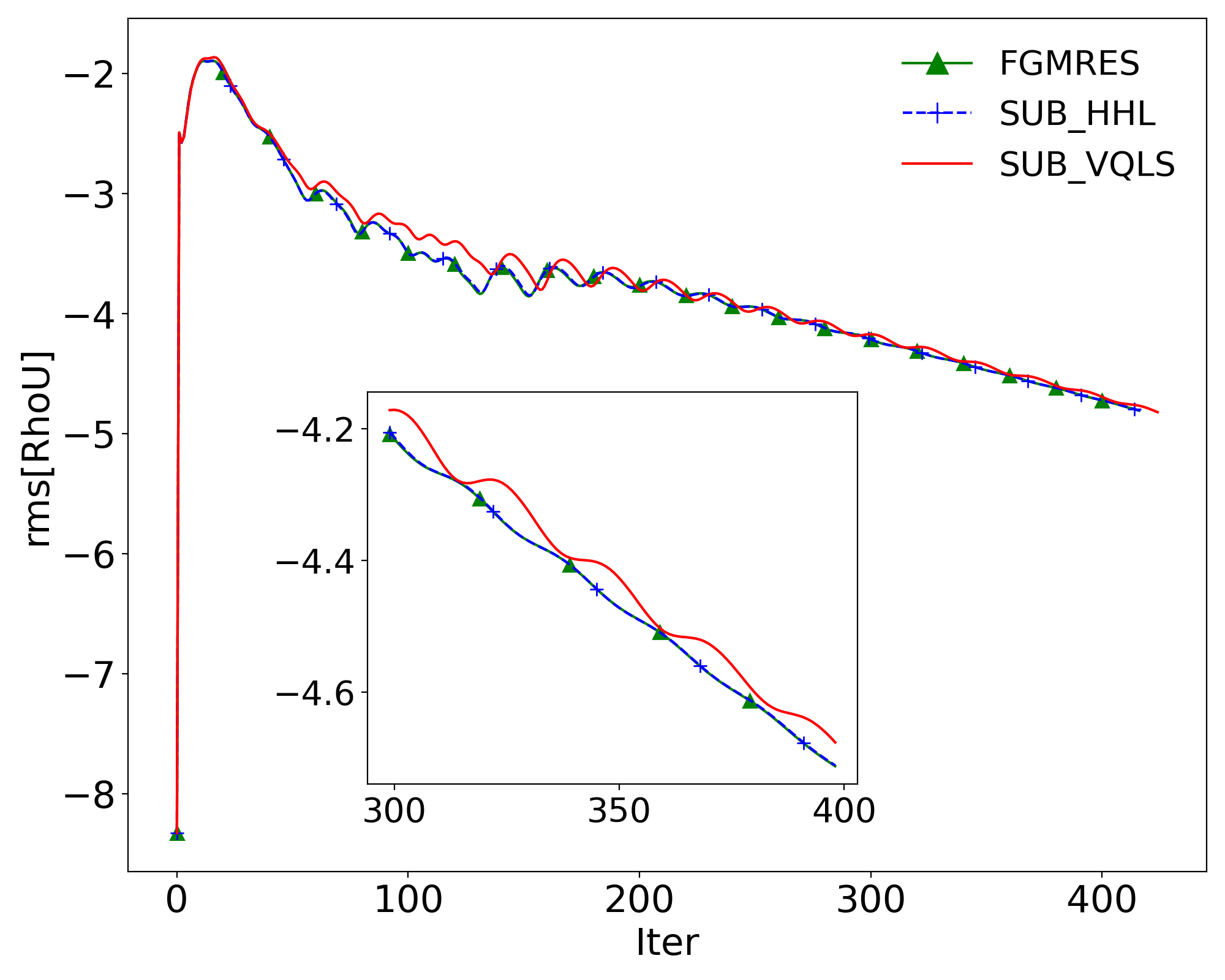}
               \put(3,76){(b)}
            \end{overpic}
        \end{minipage}
    }
    \\
    \subfigure {\
    \begin{minipage}[b]{0.45\textwidth}
            \centering
               \begin{overpic}[scale=0.315]{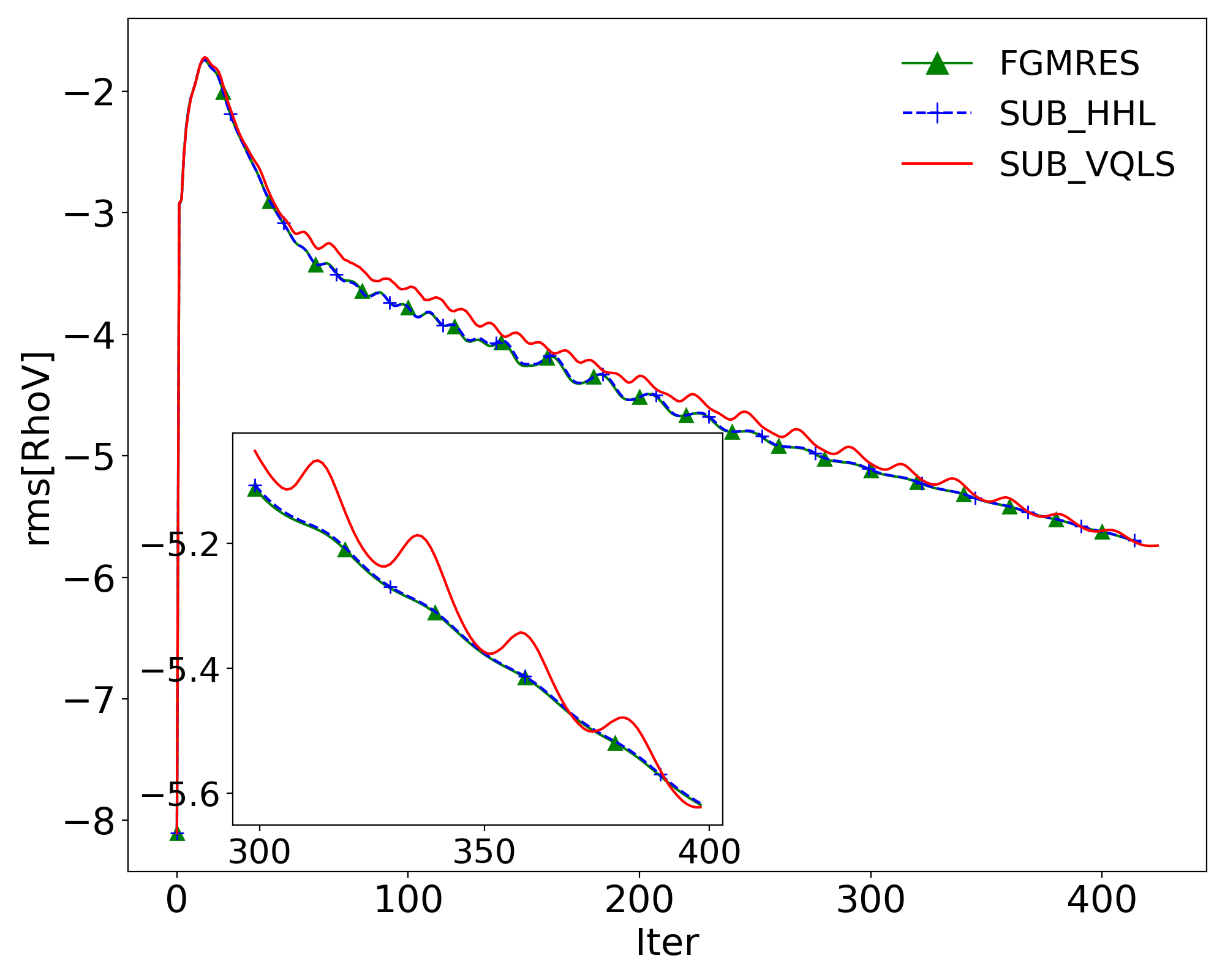}
               \put(3,76){(c)}
            \end{overpic}
        \end{minipage}
    }
    \subfigure {\
    \begin{minipage}[b]{0.45\textwidth}
            \centering
               \begin{overpic}[scale=0.315]{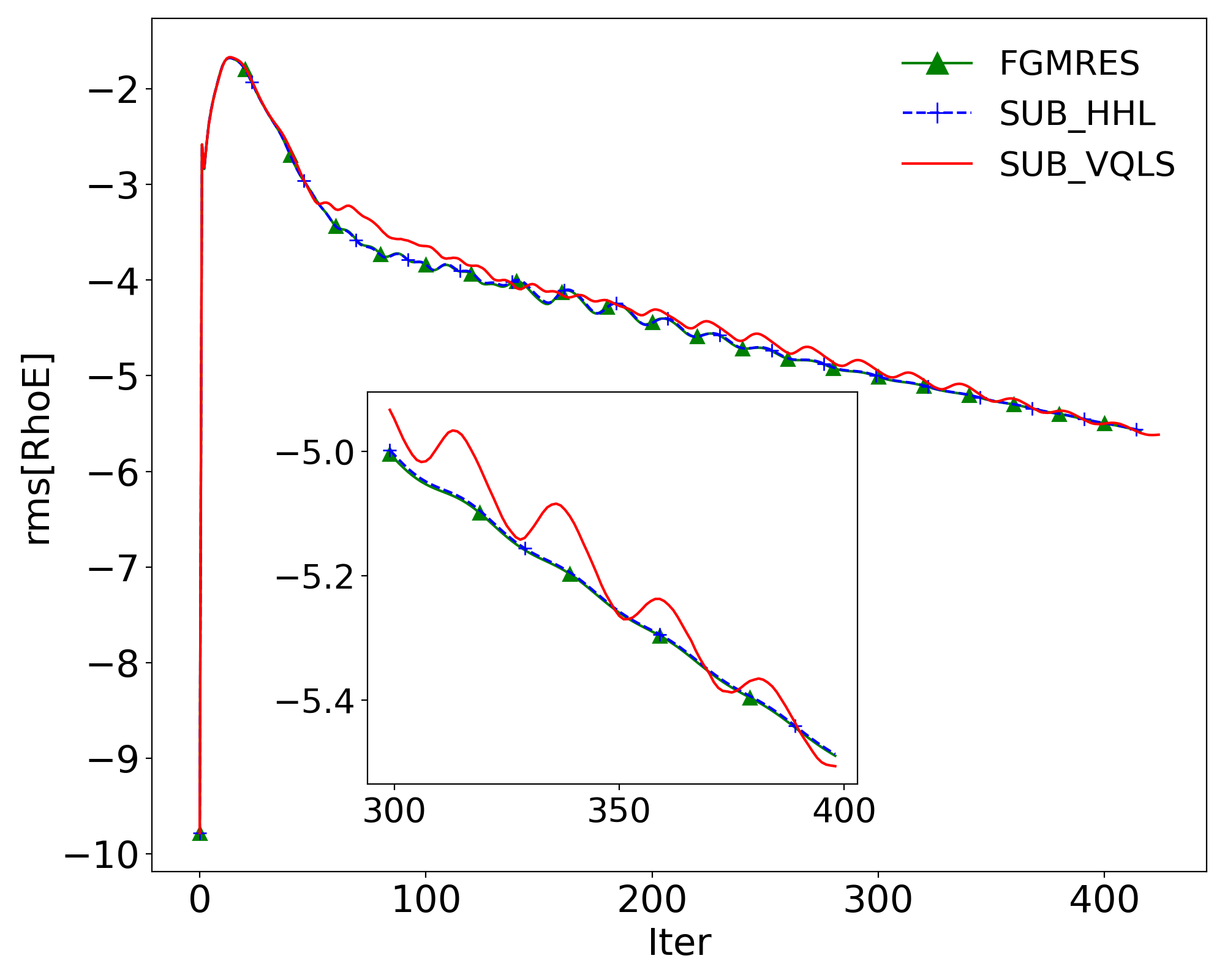}
               \put(3,76){(d)}
            \end{overpic}
        \end{minipage}
    }
    \caption{Convergence histories of residuals. The green solid lines with triangles represent data of the classical FGMRES method, while the blue dashed lines with cross markers for SUB\_HHL, and the red solid lines for SUB\_VQLS.}
    \label{fig:CHN-F1-rms}
\end{figure*}

The convergence histories of each equation are shown in Figure \ref{fig:CHN-F1-rms}. It can be seen that the CFD solver integrated with all classical and quantum linear solvers converge in a very similar trend. For quantum resource requirements, the HHL method needs 10 qubits on average, while the VQLS method needs only 4 qubits. 

\begin{figure*}[tbph]
\centering
\subfigure {\
\begin{minipage}[b]{0.42\textwidth}
        \centering
            \begin{overpic}[scale=0.34]{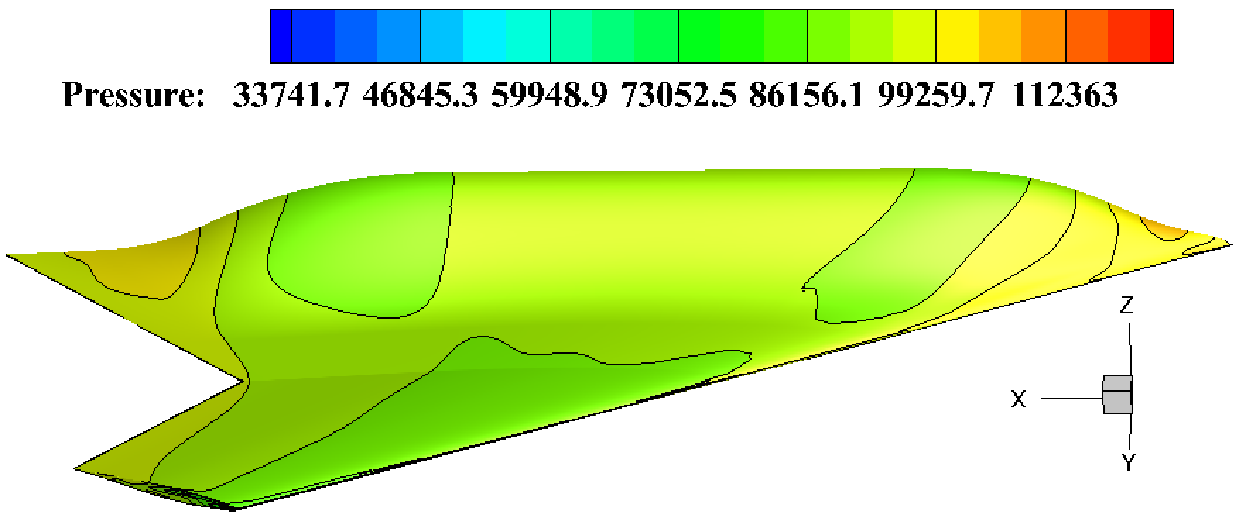}
            \put(3,38){(a)}
            \put(3,27){\fontsize{8pt}{12pt}\selectfont SUB\_HHL}
        \end{overpic}
    \end{minipage}
}
\subfigure {\
\begin{minipage}[b]{0.42\textwidth}
        \centering
           \begin{overpic}[scale=0.34]{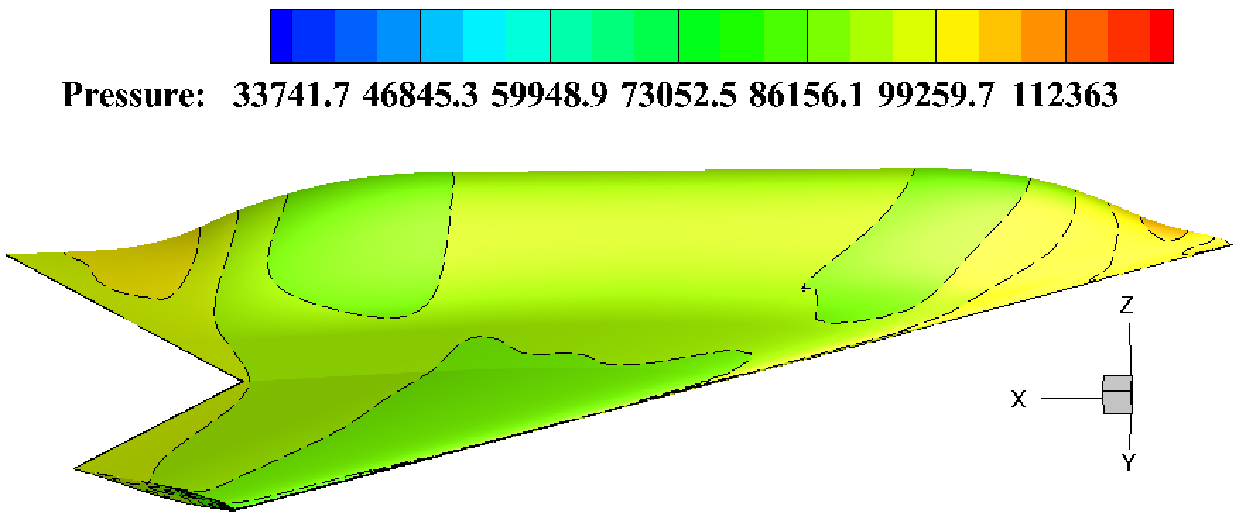}
           \put(3,38){(b)}
           \put(3,27){\fontsize{8pt}{12pt}\selectfont SUB\_VQLS}
        \end{overpic}
    \end{minipage}
}
\caption{ The pressure distribution comparison for classical and quantum solvers(SUB\_HHL and SUB\_VQLS). The flood distribution is the result of classical solver FGMRES, and the contour lines represent the quantum results. (a) Comparison result of classical FGMRES method and the quantum SUB\_HHL method.  (b) Comparison result of classical FGMRES method and the quantum SUB\_VQLS method. }
\label{FigCHN-F1-pressure}
\end{figure*}

\begin{figure}[tbph]
\centering
\subfigure {\
\begin{minipage}[b]{0.42\textwidth}
        \centering
            \begin{overpic}[scale=0.36]{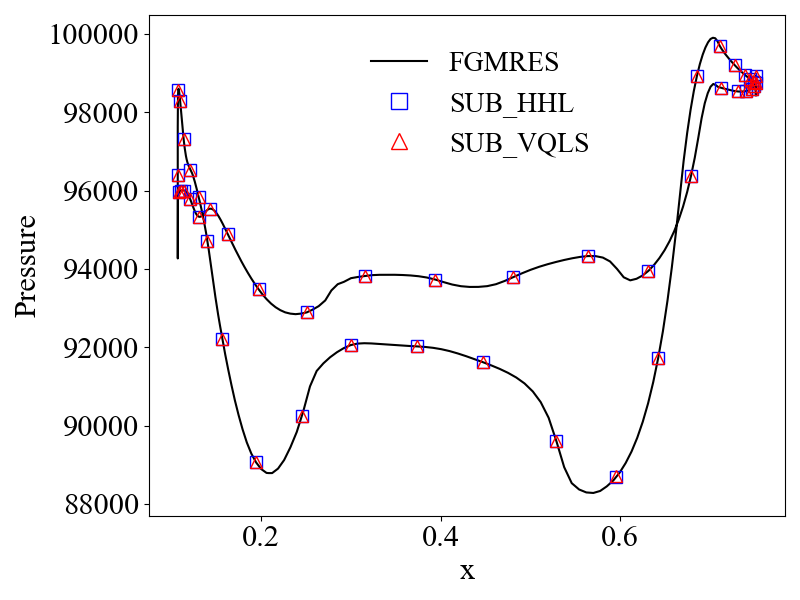}
            \put(3,73){(a)}
        \end{overpic}
    \end{minipage}
}
\\
\subfigure {\
\begin{minipage}[b]{0.42\textwidth}
        \centering
           \begin{overpic}[scale=0.36]{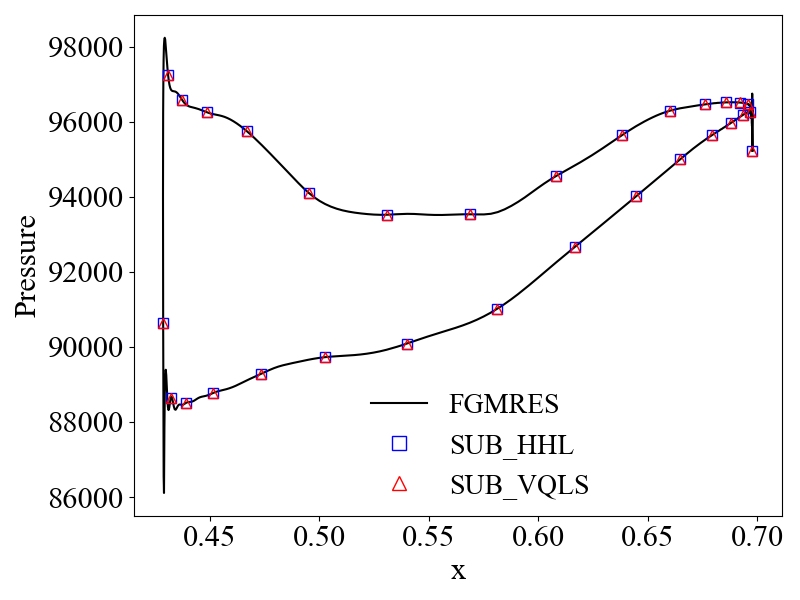}
           \put(3,73){(b)}
        \end{overpic}
    \end{minipage}
}
\caption{ The density distribution comparison for classical and quantum solvers at cross-section y= 0.05 and 0.2.  }
\label{FigCHN-F1-line}
\end{figure}

\begin{figure}
		\centering								
 	\resizebox{0.45\textwidth}{!}{\includegraphics{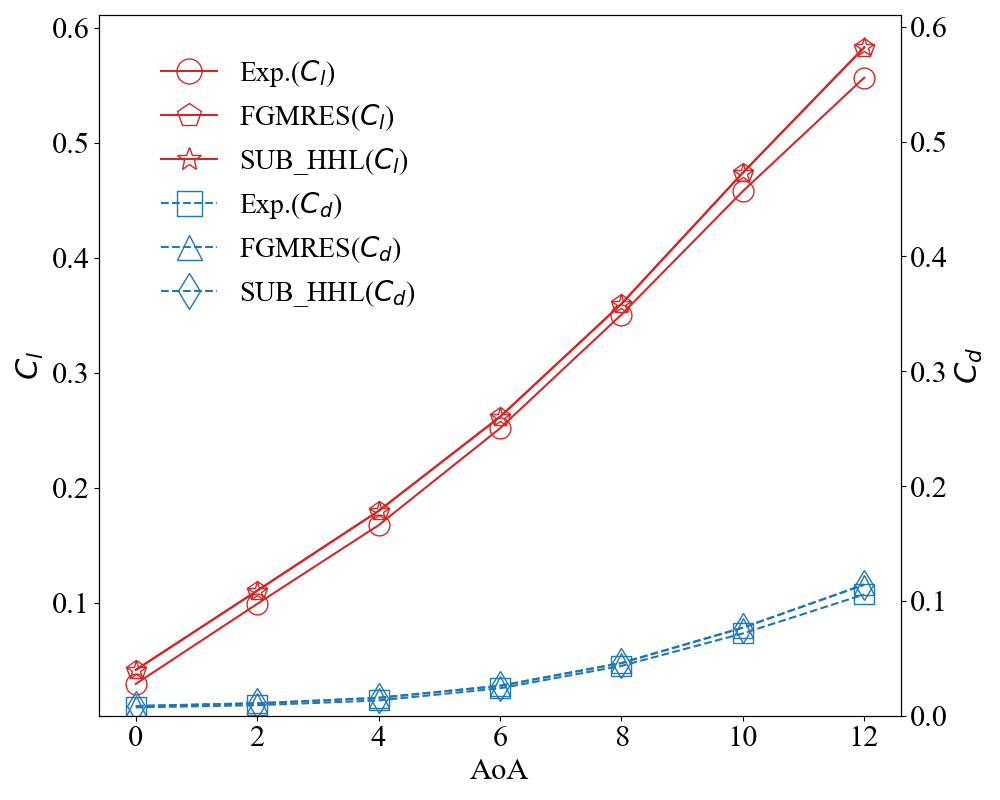}	} 
		\caption{Comparison of lift coefficient $C_l$ and drag coefficient $C_d$ obtained using classical methods, hybrid quantum-classical methods, and experimental data at different angles of attack(AoA). The solid red line represents $C_l$, while the dashed blue line for $C_d$. Red circles represent experimental data for $C_l$, while red pentagons FGMRES, and red five-pointed stars for SUB\_HHL. The blue squares represent experimental data for $C_d$, while the blue triangles for  FGMRES, and the blue diamonds for SUB\_HHL.}	
		\label{FigExp}					
\end{figure}

Besides, Figure \ref{FigCHN-F1-pressure} shows a pressure distribution comparison. For a more accurate comparison, Figure~\ref{FigCHN-F1-line} shows distribution comparison at cross sections at y= 0.05 and 0.2. It is obvious that contours computed with quantum algorithms completely coincide with those of the classical algorithms. Figure~\ref{FigExp} shows the comparison results of the overall drag and lift coefficients of cases with different AoAs. The numerical results are also compared with the experiment results. For numerical simulations, the quantum results coincide with the classical results again. Moreover, good agreement is observed for numerical results and experiment results. This case indicates that the quantum algorithms and the proposed hybrid quantum-classical CFD framework are valid in practical engineering applications.

\section{Conclusion} \label{ConclusionandOutlook}
This work exploits the potential of quantum computing in large-scale scientific computing and provides a feasible strategy to utilize quantum computing in solving CFD problems in the foreseeable future. In the current framework, nonlinear fluid problems are converted to linear problems that can be solved with quantum linear solvers. By using a quantum linear solver proposed in our previous work, this framework breaks through the limitation of the problem size limitation on certain quantum resources and provides high-precision solutions for linear equations. This quantum CFD framework can be easily integrated into many current CFD solvers and is compatible with future quantum computers. Several typical cases and a practical engineering case are computed, and the quantum results show good agreement with the classical results, which validates the feasibility of the hybrid quantum-classical CFD framework and the reliability of the quantum algorithms. 

\section*{Acknowledgments}
This work is supported by the National Key Research and Development Program of China (Grant No. 2023YFB4502500) and the Aeronautical Science Foundation of China (Grant No. 2022Z073004001).



\bibliographystyle{elsarticle-num} 
\bibliography{ref}
\end{document}